# Spatiotemporal programming via asymmetric dielectric engineering for non-volatile 2D optoelectronics

Xiaoguang Luo[#,*], Xiaolong Zhang[#], Honglei Chen[#], Jiaming Wang[#], Fan Liu, Jiongtao Zhang, Junqiang Zhang, Yihan Yin, Jinpeng Xu, Lei Ying, Renjing Xu[*], Yingchun Cheng[*], Xuetao Gan[*], and Wei Huang[*]

**ABSTRACT:** Ambipolar two-dimensional (2D) semiconductors integrated with floating-gate architectures offer a promising platform for non–volatile, reconfigurable electronics. However, the switching between p–n and n–p junction polarities has conventionally required complex multi–gate designs, hindering the scalability and integration density. Here, we demonstrate a spatiotemporal programming strategy using a dual–floating–gate architecture with a symmetry-broken tunneling dielectric. An asymmetric dielectric stack creates distinct tunneling thresholds for two floating gates, enabling a single input gate to encode spatial doping profiles in the 2D channel via defined voltage pulse sequences. We achieve on-demand, non–volatile, and reversible switching between p–n and n–p configurations with excellent retention and endurance. The reconfigurable homojunction serves as a multifunctional platform for logic encoding, rectification, photodetection, and in-sensor computing. This work establishes a design paradigm that replaces spatial input complexity with spatiotemporal programming, paving the way for high–density, multifunctional intelligent hardware.

**KEYWORDS:** Two-dimensional homojunction, spatiotemporal modulation, dual-floating gate, reconfigurable, non-volatile, photodetector

The rapid advancement of artificial intelligence and edge computing demands hardware that transcends static functionality [1-4]. Current systems, built on the von Neumann architecture, rely on discrete units for sensing, memory, and processing, suffer from fundamental bottlenecks in speed, energy efficiency, and insufficient storage space and communication bandwidth [5-7], particularly for data–intensive tasks like real–time vision [8]. A promising solution is the in–memory sensing and computing (IMSC) paradigm, which calls for a single, reconfigurable hardware element capable of dynamically encoding multiple electronic and optoelectronic functions on demand [9-12]. Such IMSC system, however, is challenging in conventional complementary metal–oxide–semiconductor image sensors, which rely on chemically doped silicon photodiodes with non–reconfigurable photoresponse [13, 14].

To address this issue, researchers have developed non–volatile, reconfigurable optoelectronic devices based on atomically thin two–dimensional (2D) semiconductors [15, 16]. The weak electrostatic screening allows 2D channels to be reconfigured into different-polarity homojunctions via electrostatic or interfacial gating [17-20]. Non–volatility in such systems is typically achieved through floating–gate charge trapping [21, 22], ferroelectric polarization [20, 23-25], and ionic/vacancy migration [26, 27]. While promising, ferroelectric and ionic approaches often suffer challenges related to material and interface quality, retention time, stability and optical transparency. In contrast, the floating–gate approach benefits from excellent retention and endurance, well–understood operating principle, and straightforward integration. Recent work has demonstrated nanosecond write/erase speeds in hexagonal boron nitride/graphene (hBN/Gr) floating–gate memories [28, 29]. However, existing floating–gate–based implementation of 2D homojunction reconfiguration require two independent floating gates that programmed spatially by separate gate electrodes [30-32], a complex architecture that impedes scalability and broader application.

Here we present a non–volatile, reconfigurable 2D optoelectronic device based on a dual–floating gate (DFG) architecture with asymmetric dielectric stack. The device is programmed by a spatiotemporal modulation via a single input gate: a defined sequence of gate–voltage pulses writes a stable, non-volatile spatial doping profile into the 2D semiconductor channel, enabling the on–demand formation of p–n or n–p homojunction. This conversion is governed by thickness–dependent Fowler–Nordheim (FN) tunneling. We demonstrate that this single non–volatile, reconfigurable element can be programmed as a logic encoder, a rectifier, and a photodetector. Furthermore, we exploit its programmable responsivity as a dynamic synaptic weight to realize in–sensor computing, implementing conceptually both hardware–based convolutional image processing and a physical artificial neural network (ANN) for action recognition. Validated with $MoTe_2$ and $WSe_2$, our DFG architecture provides a general and versatile platform that merges sensing, memory, and processing, establishing a promising paradigm for efficient, multifunctional hardware in intelligent systems.

Fig. 1a illustrates the concept of the presented reconfigurable DFG optoelectronic device, fabricated via 2D van der Waals integration. The structure comprises two split Gr flakes separated by a micrometer–scale gap on $SiO_2$/Si substrate, which function as the dual–floating gate. A first hBN flake is aligned with the gap on one graphene flake, while a second larger one encapsulates two graphene flakes both, forming the thick and thin dielectrics above two floating gates. A few–layer $MoTe_2$ flake across the gap works as the active channel (Fig. S1). Channel polarity is determined by charges trapped in the Gr floating gates that programmed via tunneling by applying a gate voltage to the Si substrate. The floating–gating behavior enables non–volatile control, and its programmed state is retained for a long time after removing the gate voltage. Generally, thick dielectric layer weakens the charge tunneling, resulting in larger threshold voltage for programming with increased dielectric thickness. Therefore, applying gate–voltage pulses with specified amplitudes and polarities will reconfigure the $MoTe_2$ channel into p–n or n–p homojunction (Fig. 1a). For example, a large positive gate–voltage pulse followed by a small negative pulse creates a p–n homojunction within the $MoTe_2$ channel, whereas the reverse sequence (a large negative pulse followed by a small positive one) forms an n–p homojunction.

In the hBN/Gr single–floating gate architecture, charge trapping in the Gr floating gate saturates more rapidly with higher control gate voltages and thinner hBN dielectrics (see Supporting Information S4), which even enable writing/erasing speed as fast as 20 ns [28, 33]. To ensure stable device operation, a gate–voltage pulse duration of 1 s was set in this study, sufficient for full charge saturation in the Gr floating gate. The tunneling current to Gr floating gate can be calculated theoretically using a physics–based model with WKB approximation [34]. A typical device shown in Fig. 1b is employed to illustrate electrical and optoelectronic reconfiguration. Fig. 1c shows two output characteristic curves ($I_{DS}$ vs. $V_{DS}$) following the application of gate–voltage dipulses of (60, –30) and (–70, 21) V, respectively. The distinctly opposite rectification behaviors, both with a ratio of ~$10^4$, confirm the polarity reconfiguration of the device junction. According to the definition in Fig. 1a, the output curve for the (60, –30) V dipulse corresponds to a p–n homojunction, while that for the (–70, 21) V dipulse corresponds to an n–p homojunction. Accounting for the series resistance of $MoTe_2$ channel, the current is well described by a modified Shockley equation incorporating the Lambert $W$–function [35]. The fitted ideality factors for the p–n and n–p cases are 1.2 and 1.5, respectively. For these junctions, the ideality factor of 1 indicates carrier transport dominated by diffusion, whereas the factor of 2 suggests a recombination–dominated process. In our device, the output characteristics—and thus the ideality factor—are modulated by the gate–voltage dipulse (with the ideality factor varying between 1 and 2, Fig. S6). This tunability arises from dipulse–controlled Fermi level of the 2D channel, which subsequently alters the depletion of the junction [36].

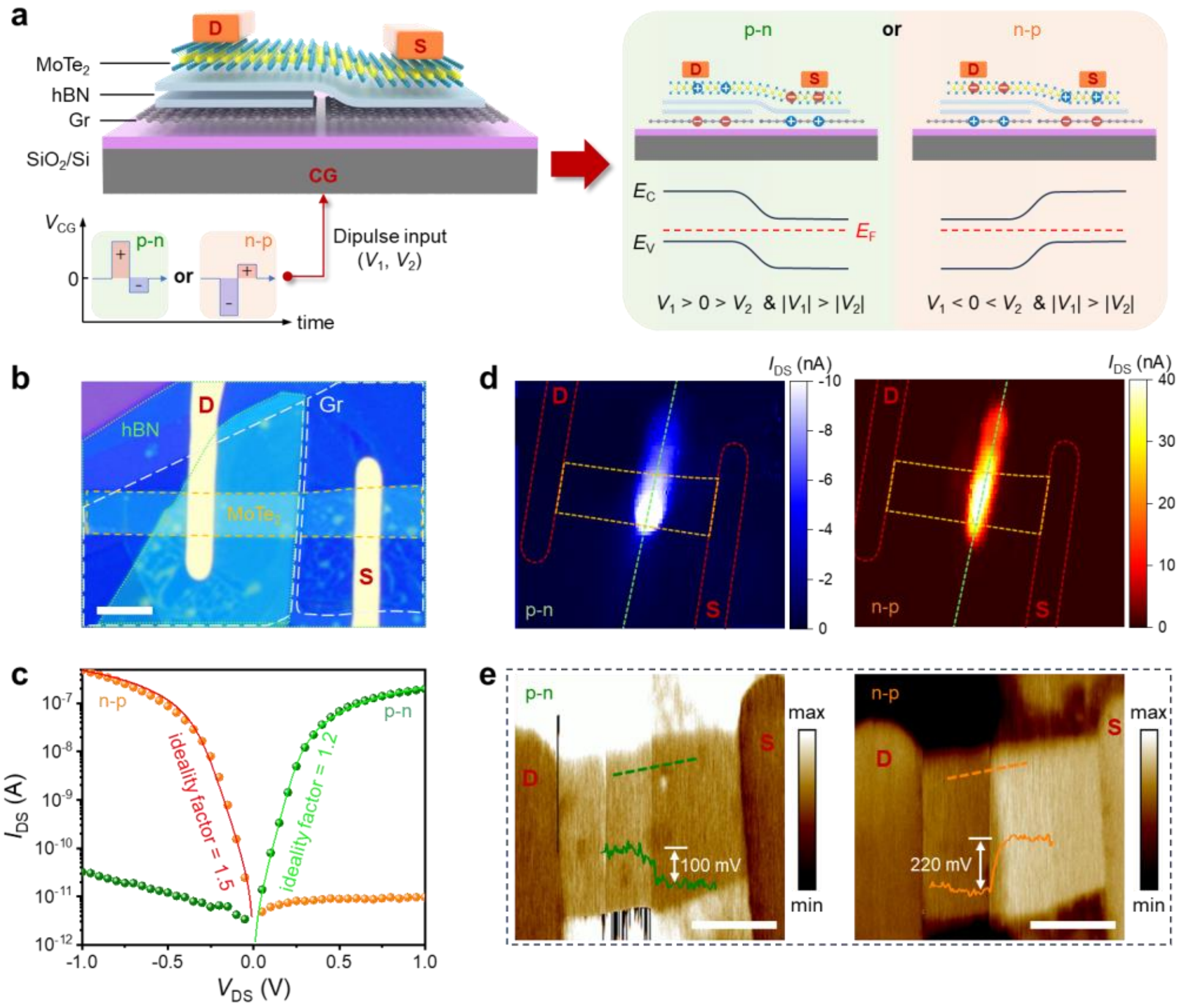

**Fig. 1. Device characterizations of $MoTe_2$ photodetectors with dual-floating gate. a)** Schematic diagram (left) of the $MoTe_2$ DFG photodetector and band structure (right) of the $MoTe_2$ homojunction at the spatiotemporal–programmed p–n and n–p configurations. D: drain, S: source, and CG: control gate. **b)** Optical image of a device. Scalebar: 10 μm, Gr: graphene. **c)** Output characteristics of the device at p–n ($V_1 = 60$ V, $V_2 = -30$ V) and n–p ($V_1 = -70$ V, $V_2 = 21$ V) configurations, respectively. **d)** Photocurrent mapping of the device at $V_{DS} = 0$ V for p–n ($V_1 = 60$ V, $V_2 = -25$ V) and n–p ($V_1 = -60$ V, $V_2 = 24$ V) configurations under activated by the focused 532 nm laser, with the laser spot diameter ~3 μm and laser power ~32 μW. **e)** KPFM characterization of a device at p–n ($V_1 = 70$ V, $V_2 = -18$ V) and n–p ($V_1 = -70$ V, $V_2 = 20$ V) configurations, respectively. Insets: profiles of contact potential difference along the dashed lines. Scalebar: 5 μm.

To verify that the tunable rectification originates from the reconfiguration of the $MoTe_2$ homojunction, we performed spatially resolved scanning photocurrent microscopy at $V_{DS} = 0$ V under 532 nm illumination for both p–n and n–p configurations. The photocurrent map in Fig. 1d confirms that the dominant photocurrent for both configurations is produced at the $MoTe_2$ homojunction, rather than at the Au/$MoTe_2$ contacts or within the uniform $MoTe_2$ channel. In the homojunction, the photo–generated electron–hole pairs are separated by the built–in electric field, and the resulting carriers diffuse to respective electrodes to generate the measured photocurrent. The observed negative and positive photocurrents for p–n and n–p configurations, respectively, stem from the opposite directions of the built–in electric field, directly demonstrating the homojunction reconfiguration. This reconfiguration was further visualized via Kelvin probe force microscopy (KPFM), which maps the work function distribution of the device (Supporting Information S3 and Fig. 1e). For the given programmed gate–voltage dipulse, the work function on left side of the $MoTe_2$ homojunction is 100 meV larger than on the right in the p–n configuration, whereas in the n–p configuration, the right–side work function is 220 meV larger than the left. This clear inversion in the work function profile provides direct evidence of the opposite band bending, confirming the successful reconfiguration of the homojunction.

The reconfiguration arises from the floating–gate effect (driven by gate–voltage dipulse applied to the control gate), which is also evident in the devices' transfer characteristics ($I_{\mathrm{DS}}$ vs. $V_{\mathrm{CG}}$). Typical dual–sweep transfer curves for a $MoTe_2$/hBN/Gr DFG device (Figs. 2a, S7a and S7b) exhibit two distinct memory windows, denoted by $\Delta V_{\mathrm{th}}^{+}$ and $\Delta V_{\mathrm{th}}^{-}$, in the p–branch and n–branch of the ambipolar transfer curves, respectively. These windows result from the tunneling and trapping of holes and electrons in the Gr floating gates. Beyond a threshold, the memory windows broaden linearly with the gate–voltage sweep range. The consistent observation that $\Delta V_{\mathrm{th}}^{-} > \Delta V_{\mathrm{th}}^{+}$ indicates a higher electron trapping efficiency in this device. In contrast, the pronounced asymmetric windows are absent in a $MoTe_2$/hBN/Gr single–floating gate device (Fig. S7d), ruling out the difference in barrier heights at the $MoTe_2$/hBN and hBN/Gr interfaces as the primary cause. We also attribute the asymmetry to non–uniform tunneling paths [30, 37], a phenomenon commonly observed in the asymmetric floating–gate architectures [21, 22]. When the gate–voltage is directly applied to both two Gr floating gates, the memory windows vanish dramatically in the transfer curves (Figs. S8a-S8c), confirming the floating-gate role of Gr flakes. The underlying tunneling mechanisms were investigated by the current–voltage characteristics between D/S electrodes and Gr floating gates, i.e., the dielectric layer composed of double/single hBN flakes, respectively. These measurements (Figs. S8d and S8e) reveal two distinct regimes: direct tunneling at lower bias and FN tunneling at higher bias. The current in the FN tunneling regime is several orders of magnitude larger. Therefore, the charge injection into or extraction from the floating gates occurs primarily through FN tunneling.

A transition voltage ($V_{\mathrm{T}}$) marks the switch between two tunneling regimes, which is higher for the double–hBN–flake dielectric. This transition voltage exhibits a direct, physical dependence on the total dielectric thickness, regardless of whether the dielectric comprises single or double hBN flakes. We confirmed this relationship through measurements on 12 $MoTe_2$/hBN/Gr heterostructures with varying total thicknesses of hBN dielectrics (Fig. S9). By applying a bias voltage to the drain electrode (Fig. 2b), we extracted positive and negative $V_{\mathrm{T}}$ values, signifying FN tunneling regimes for both charge injection and extraction. As summarized in Fig. 2c, $V_{\mathrm{T}}$ values indeed exhibit a positive correlation with the total hBN thicknesses.

In a $MoTe_2$/hBN/Gr DFG device, the operating gate voltage for switching the charge states of Gr floating gates is governed by the hBN thickness, e.g., $|V_{\mathrm{T}}^{\mathrm{s}}| < |V_{\mathrm{T}}^{\mathrm{d}}|$ with $V_{\mathrm{T}}^{\mathrm{s/d}}$ denoting the transition gate voltage for single/double–hBN–flake dielectric. This asymmetry enables junction reconfiguration. An initial gate–voltage pulse with amplitude $|V_1| > |V_{\mathrm{T}}^{\mathrm{d}}|$ charges both Gr floating gates with the same kind of charge carriers, and a subsequent pulse with amplitude $|V_{\mathrm{T}}^{\mathrm{s}}| < |V_2| < |V_{\mathrm{T}}^{\mathrm{d}}|$ then selectively discharges only the Gr floating gate under single–hBN–flake dielectric. When $V_1$ and $V_2$ are of opposite polarity, two Gr floating gates trap electrons and holes, respectively, programming the $MoTe_2$ channel into p–n or n–p homojunction. Fig. 2d shows the continuous regulation of the homojunction with varying $V_2$ when $|V_1| = 70$ V. For $V_1 = 70$ V and a negative $V_2$, both Gr floating gates trap electrons after the initial pulse, and then the charge state of the Gr floating gate under single–hBN–flake dielectric is progressively switched from electrons to holes as the amplitude of $V_2$ increases, resulting in the transition of $MoTe_2$ homojunction from p–p to p–n. An analogous n–n to n–p transition occurs for $V_1 = -70$ V with an increasing positive $V_2$. The $MoTe_2$ homojunction is obviously reconfigured into p–n and n–p types by gate-voltage dipulses of (70, –20) and (–70, 20) V, respectively. The corresponding electron dynamics, driven by FN tunneling between Gr floating gates and $MoTe_2$ channel, are illustrated in the band diagrams of Figs. 2e and 2f. This programming scheme effectively converts a defined temporal sequence of gate–voltage pulses into a stable, non–volatile spatial doping profile within the semiconductor channel. Following this spatiotemporal modulation, the two floating gates at distinct charge states independently gate the local regions of semiconductor channel above them.

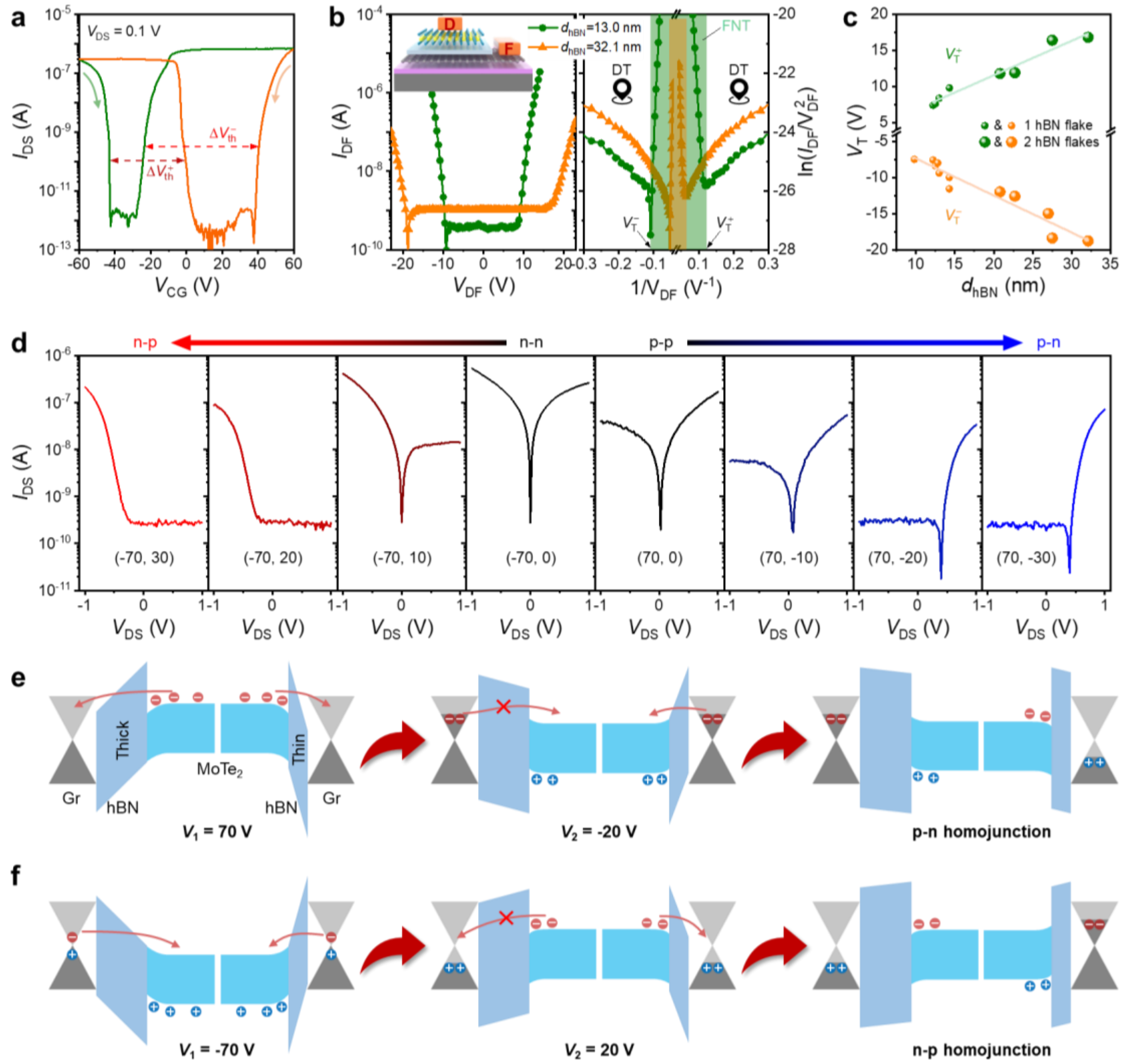

**Fig. 2. Reconfigurable $MoTe_2$ homojunctions defined by dual-floating gate. a)** Dual–sweep transfer characteristics of the device at $V_{DS} = 0.1$ V, where $\Delta V_{th}^{+}$ and $\Delta V_{th}^{-}$ denote the memory windows at p–branch and n–branch, respectively. **b)** Tunneling current ($I_{DF}$) through the $MoTe_2$/hBN/Gr structure with the dielectric composed of either single hBN flake (green) or double hBN flakes (orange). Left panel: $I_{DF}$ with respect to applied voltage ($V_{DF}$), right panel: the corresponding FN plots. Inset: schematic of the measured structure. For the single hBN flake case, the FN tunneling regime (green region) and direct tunneling regime (white region) are separated by the transition voltages of $V_{T}^{\pm}$. DT: direct tunneling, FNT: FN tunneling. **c)** Transition voltages as a function of hBN thickness, where the solid lines are guides for the eye. **d)** Reconfiguration of the homojunction type via gate–voltage dipulses. **e-f)** Band diagrams illustrating the formation of p–n and n–p homojunctions under (70, -20) and (-70, 20) V dipulses, respectively, driven by the FN tunneling of electrons.

Fig. 3a demonstrates the switching between low–current (high–resistance $R_H$) and high–current (low–resistance $R_L$) states of a device that programmed into p–n and n–p configurations by (70, –30) and (–70, 30) V dipulses, respectively. The programmed states remain stable during the 100 s transient current measurements at $V_{DS} = \pm 1$ V, with the on–off ratios of approximately 2000 and 600, respectively. Excellent retention is evidenced by the stable performance of a device fabricated 7 months ago (Fig. S10a), although a measurable degeneration onset emerges after several hundred seconds. The endurance characteristics of a $MoTe_2$/hBN/Gr DFG device are presented in Figs. S10b and S10c, where the on-off ratio exceeding $10^4$ are maintained over 600 cycles at both $V_{DS} = \pm 1$ V. These combined retention and endurance results suggest high reliability of our spatiotemporal programming strategy. Based on the non–volatile reconfiguration, the device functions as a diode with programmable rectification. The rectification ratios of the device (Fig. 3a) at $|V_{DS}| = 1$ V are approximately 1800 for the p–n configuration ($R_H \sim 3.0 \times 10^9$ Ω, $R_L \sim 1.7 \times 10^6$ Ω) and 700 for the n–p configuration (with $R_H \sim 3.5 \times 10^9$ Ω, $R_L \sim 5 \times 10^6$ Ω). When connected in series with a resistor of value $\sim 10^{[\log(R_H)+\log(R_L)]/2}$, the device then can serve as a signal filter or encoder. Fig. 3b demonstrates this switchable filtering operation using a 100 MΩ resistor and a sinusoidal voltage input. When programmed to p–n configuration, the circuit transmits the positive voltage component, whereas in the n–p configuration, it transmits the negative component. This rectification functionality can be extended by replacing the passive resistor with a second, back–to–back $MoTe_2$/hBN/Gr DFG device, which operates analogously to a CMOS–like device.

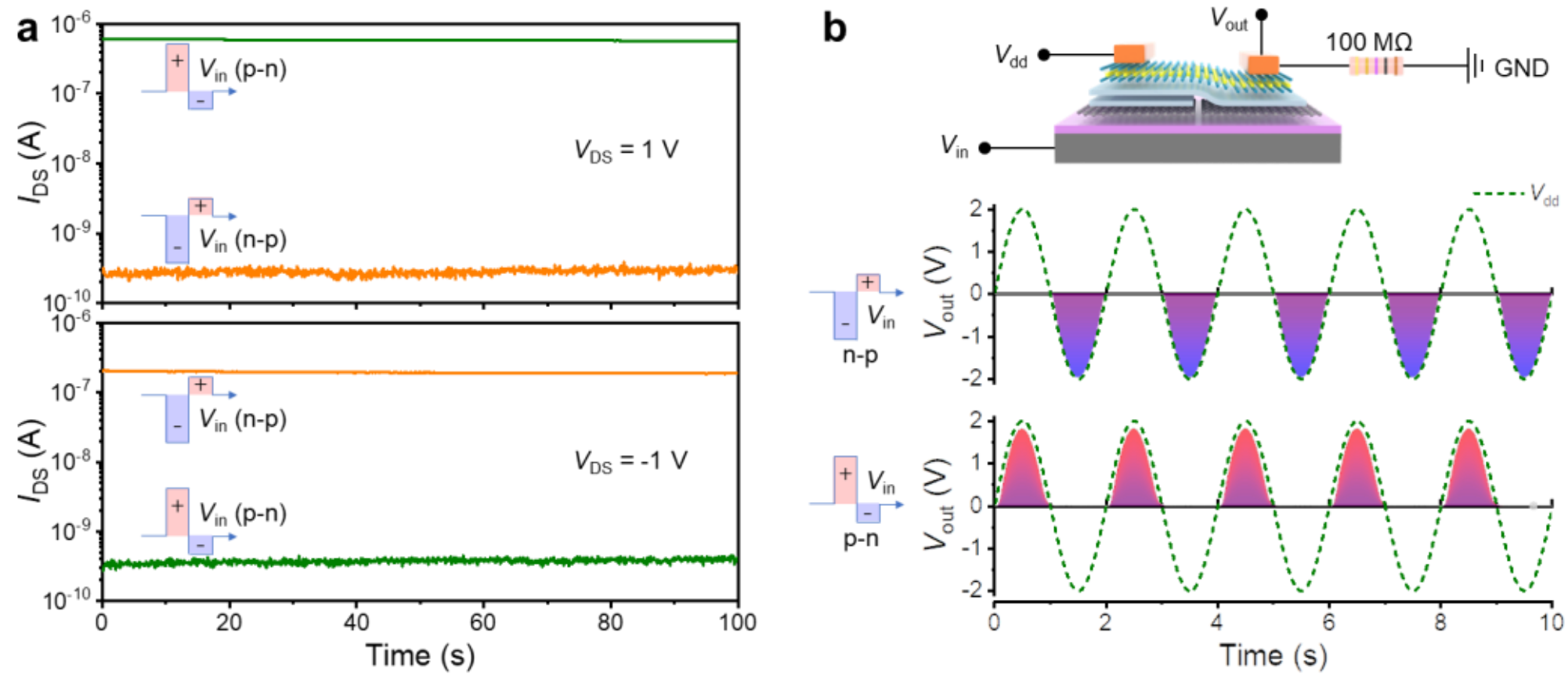

**Fig. 3. Logic encoding and reconfigurable rectification and. a)** Retention performance of the DFG device programmed into p–n (70 V, -30 V) and n–p (-70 V, 30 V) states, measured at $V_{DS} = 1$ V (upper) and $V_{DS} = -1$ V (lower). **b)** Programmable rectification of a sinusoidal signal for n–p (-70 V, 16 V) and p–n (70 V, -14 V) states, with the device in series with a 100 MΩ load.

The non–volatile reconfiguration of the homojunction enables a programmable photoresponse. Under 532 nm illumination, the output curves exhibit a zero–bias shift for both p–n and n–p configurations (Fig. S11). The shift directions refer to voltage are opposite between two configurations, consistent with the dynamic behavior of photovoltaic carriers in conventional p–n and n–p junctions [38-40]. For the p–n configuration, the photovoltaic output yields a large open-circuit voltage of 0.47 V and a high power conversion efficiency of 3.3%. To quantify the photodetection performance at photovoltaic mode, transient photocurrent when $V_{DS} = 0$ V was measured at different laser power density ($P_{in}$) via a one–time test method (with the experimental setup being shown in Fig. S12a), where the laser is modulated at 1 Hz (40% duty ratio).

For both configurations, the photocurrent shows near–perfect linearity with incident laser power density (Figs. S12c and S13b). Taking the entire $MoTe_2$ channel area ($A = 194.7$ μm$^2$), the responsivity is evaluated as 44.5 mA/W for the p–n configuration. Combined with the low noise spectral density of $1\times10^{-28}$ $A^2$/Hz at low frequency, the noise equivalent power reaches to 0.23 pW·$Hz^{-1/2}$. Comparable performance metrics are obtained for the n–p configuration (Fig. S13), with the responsivity of 36.2 mA/W and the noise equivalent power of 0.28 pW·$Hz^{-1/2}$. It should be noted that the photocurrent is generated specifically within the $MoTe_2$ homojunction region (length ~ 1 μm), rather than across the entire $MoTe_2$ channel (length ~ 25 μm), as displayed in Fig. 1d. Thus, a device engineered with a channel length optimized to match the junction dimension could improve these performance metrics by at least one order of magnitude. The stability of the programmed photoresponse is confirmed by long–duration transient photocurrent measurements. For both p–n and n–p states, the photocurrent remains essentially unchanged over 500 s after a single programming event (Fig. 4a) and across 600 programming cycles (Fig. S10d), respectively.

As programmed by gate–voltage dipulse, the photoresponse states of $MoTe_2$/hBN/Gr DFG device can be encoded even under the same illumination. For a fixed initial pulse of $|V_1| = 70$ V, the photocurrent of a chosen device decreases monotonically as the amplitude of the following pulse $|V_2|$ decreases from 26 V, and already reaches zero at $|V_2| = 5$ V. This continuous tuning defines nine discrete programmable states, labeled as +4, +3, +2, +1, 0, –1, –2, –3, and –4 based on their photocurrent magnitude (with photocurrents of 0.166, 0.126, 0.084, 0.041, 0, –0.039, –0.077, –0.111, and –0.150 nA, respectively, at 14.56 mW/cm$^2$, Fig. 4b). The photoresponse time is approximately 3 ms for both p–n and n–p configurations, which could potentially be reduced by minimizing the carrier diffusion length (e.g., by shortening the $MoTe_2$ channel length) [41]. The photocurrent for each state scales nearly linearly with incident laser power density ($P_{in}$) from 0 to 14.56 mW/cm$^2$ (Fig. 4c). This direct programmability therefore encodes the device's responsivity ($R$), and the output photocurrent $I_{ph} = P \cdot R$ with $P = P_{in} \cdot A$ representing the effective laser power.

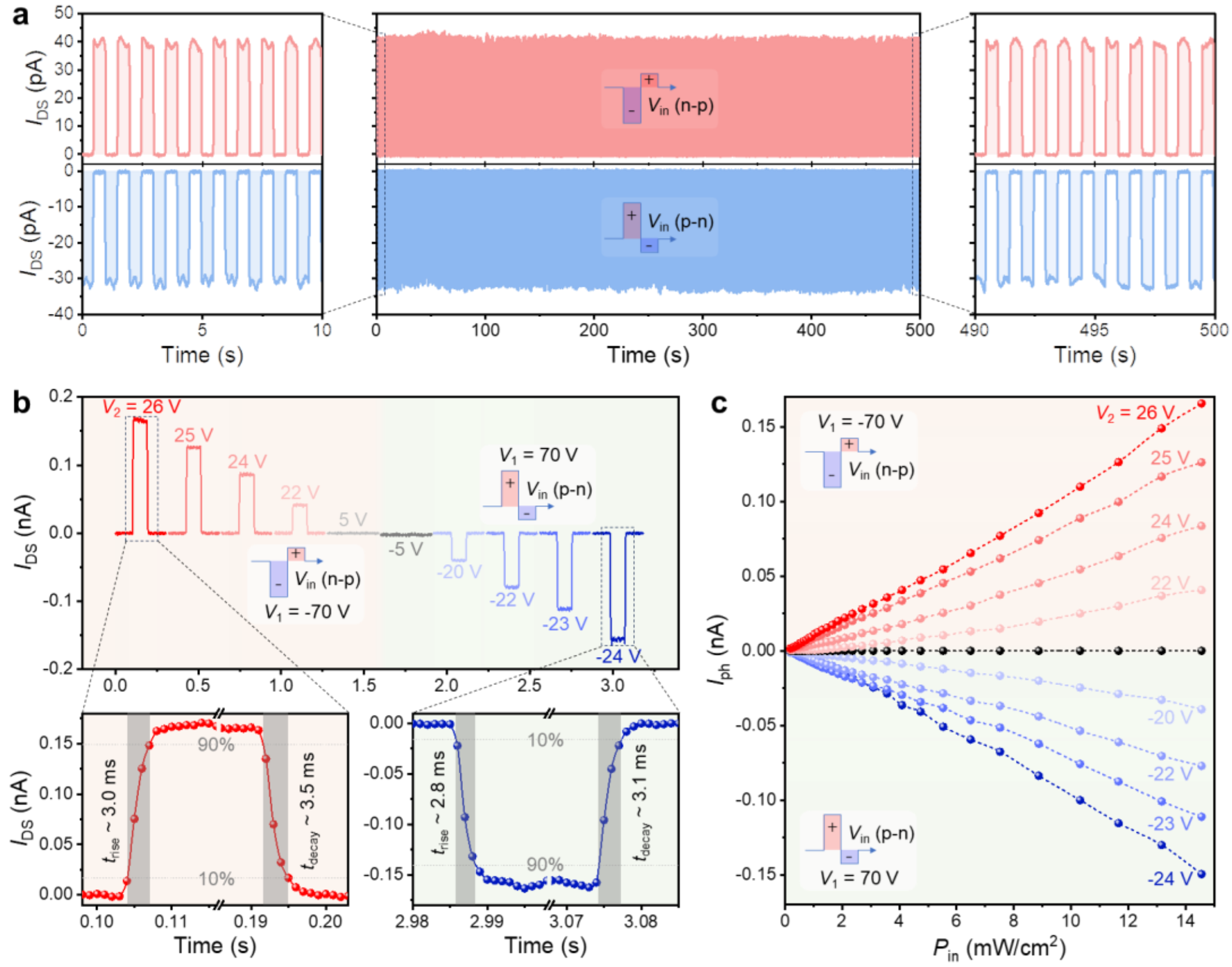

**Fig. 4. Programmable photoresponse under gate–voltage dipulses**. **a)** Transient current measured over 500 s under modulated (1 Hz) 532 nm laser illumination (1.72 mW/cm$^2$) for the n–p (-60 V, 25 V) state (upper panels) and the p–n (60 V, -25 V) state (lower panels). **b)** Transient current for ten distinct device states programmed sequentially by different gate–voltage dipulses (upper panel) under 532 nm laser illumination (14.56 mW/cm$^2$). The corresponding response times for the n–p (-70 V, 26 V) and the p–n (70 V, -24 V) states are shown in the lower panels. **c)** Photocurrent with respect to incident laser power density for nine programmable states.

With its non–volatile and reconfigurable photoresponse, the $MoTe_2$/hBN/Gr DFG device offers potential for in–sensor computing. In this scheme, the greyscale image of any object corresponds to a spatial matrix of incident optical power (**P**). The output signal matrix (**I**) read from the sensor array is given by $\mathbf{I} = \mathbf{P} \odot \mathbf{R}$, where $\odot$ denotes element-wise multiplication and $\mathbf{R}$ is the spatially programmable responsivity matrix of the array. We simulated the in–sensor convolutional processing using a 3×3 network kernel hypothetically composed of nine uniform modeled DFG devices (Fig. 5a). For each kernel position, the output photocurrent is the sum of the contributions from all nine devices, as governed by Kirchhoff's current law. The final processed image is obtained by normalizing the complete matrix of these summed photocurrents. Therefore, processing a 287×216 pixel input image yields a 285×214 pixel output. Using the experimentally characterized photoresponse from Fig. 4c, we executed in–sensor processing with four standard convolutional kernels, including Gaussian blur, image sharpen, Sobel edge detection, and embossing (kernel matrices in Fig. S14). The close correspondence between outputs from our modeled DFG device array (Fig. 5b, lower row) and conventional software simulation (Fig. 5b, upper row) validates the feasibility of the in–sensor processing based on our devices.

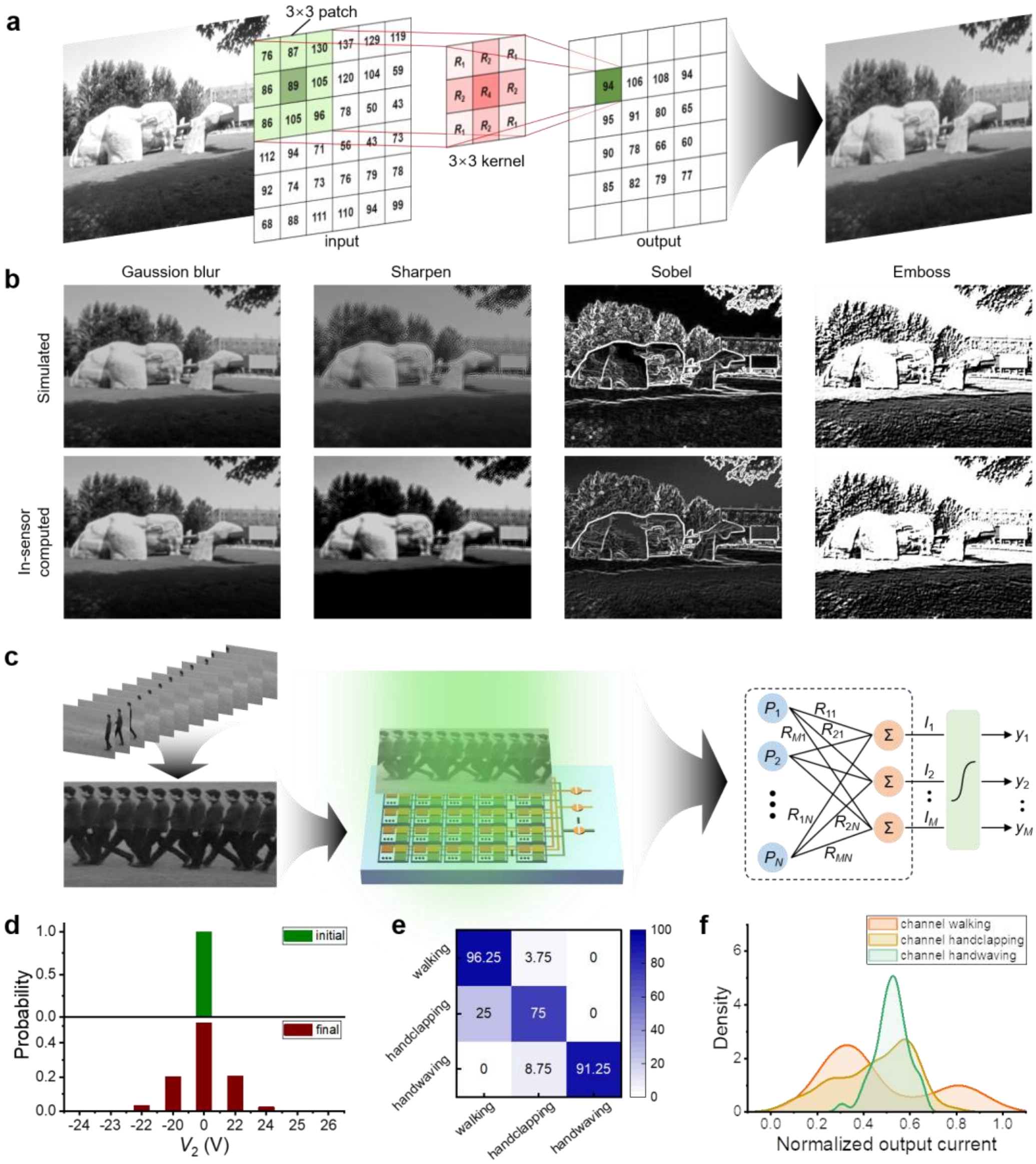

**Fig. 5. In-sensor image processing and in-sensor ANN for action classification. a)** Schematic of in–sensor processing using a 3×3 convolutional kernel of DFG devices for a 287×216 pixel input image. **b)** Output images obtained via digital processing (upper row) and in–sensor processing (lower row) for four different operations. **c)** Conceptual schematic of the in-sensor ANN computing system. **d)** Initial and final voltage distribution of the following pulse. **e)** Confusion matrix of the classification results on the test dataset. **f)** Output current distribution of the in–sensor array. The data for the three actions comes from the KTH Action Dataset[42].

Programmable responsivity functions as dynamic synaptic weight, enabling the device to form a fundamental hardware platform for in–sensor ANN computing. In a prospective hardware implementation, the crossbar architecture comprising a defined number of such devices constitutes a physical ANN, capable of executing array–scale computations in a single operational step. For classification of $M$ distinct action classes using an $N$ pixel array, we employ a strategy where each pixel integrates $M$ independently programmable subpixels. Subpixels of the same class–specific channel are interconnected, generating $M$ distinct, accumulated photocurrent outputs ($I_1$, $I_2$, $\cdots I_M$). These outputs serve as the critical parameters to an estimator for final action classification (Fig. 5c).

To conceptually demonstrate the in–sensor computing capability of the $MoTe_2$/hBN/Gr DFG device, we designed and simulated a physical ANN using $128 \times 128$ pixel array (Fig. S15), programmed with the experimentally measured photoresponse characteristics from Fig. 4c to classify three actions from the KTH Action Dataset [42]: walking, handclapping, and handwaving. The offline–trained synaptic weights are mapped to discrete photoresponse states in the simulated device array, where the gate–voltage of the following pulse follows a distribution that shown in Fig. 5d. This simulation achieves a classification accuracy of 87.50%—negligibly lower (0.42%) than full–precision software benchmarks (87.92%)—validating a high–fidelity weight transfer. The simulated spatial voltage patterns (Fig. S15b) demonstrate the array's capacity for category–specific feature extraction. A primary confusion, misclassifying "handclapping" as "walking", stems from a significant overlap in their output current distributions (Fig. 5f). This ambiguity in current domain reflects the inherent feature similarity between these two actions, and presents a fundamental challenge to the linear separability limit of our single–layer, in–sensor network architecture.

Crucially, the in–sensor ANN exhibits superior robustness against salt–and–pepper noise compared to its software counterpart (Fig. S16b). This noise immunity stems from two architectural features: (i) discrete, non–volatile physical states provide digital–like noise margins that suppress the impact of stochastic signal fluctuations, and (ii) the sparse mapped–weight distribution (Fig. 5d) functions as a mask, intrinsically insulating the network from noise propagation through inactive synaptic pathways. By performing computation directly at the sensory front–end with these robust states, our architecture avoids the error accumulation that plagues multi–layer analog processing systems. These results demonstrate that in–sensor computing with our DFG devices can deliver effective pattern recognition with high robustness, presenting a tangible pathway to mitigate the

data–movement bottlenecks and high–power consumption of conventional vision systems. A practical array–level implementation will necessitate addressing device–to–device variation, programming uniformity, and electrical interconnect effects, which may introduce weight errors and affect the accumulated photocurrent outputs. These challenges can be mitigated through array–level calibration and variation–aware weight mapping, and will be the focus of our future work on experimentally realized arrays.

In summary, we have demonstrated a non–volatile, reconfigurable optoelectronic device based on DFG featuring asymmetric tunneling dielectrics. The device's electrical and optoelectronic properties are programmed via gate–voltage dipulses, a spatiotemporal modulation of translating a temporal voltage sequence into a stable spatial doping profile. This programmability enables exceptional functional versatility, such as a stable rectifier, logic encoder, and photovoltaic photodetector. Most significantly, by using the programmable responsivities as the dynamic synaptic weights, we have performed simulation–based proof–of–concept demonstrations of in–sensor computing, including convolutional image processing and ANN–based action recognition. Our DFG architecture, validated here with $MoTe_2$ and $WSe_2$ (Fig. S17) and extensible to other ambipolar semiconductors [43, 44] or other asymmetric dielectric engineering techniques [45, 46], bypasses the traditional separation of sensor, memory, and processor, establishing a general paradigm for multifunctional, reconfigurable hardware for future integrated sensing and processing systems.

## ■ ASSOCIATED CONTENT

### Supporting Information

The Supporting Information is available free of charge at …

Detailed device fabrication, device characterization, and in–sensor ANN computing; KPFM characterizations; duration of gate voltage pulses for the devices; electrical characterizations of the devices; reconfigurable photoresponse; in-sensor image processing and action classification; universality verification (PDF)

## ■ AUTHOR INFORMATION

### Corresponding Authors

**Xiaoguang Luo** – State Key Laboratory of Flexible Electronics & School of Flexible Electronics, Northwestern Polytechnical University, Xi'an 710129, China; Email: iamxgluo@nwpu.edu.cn

**Renjing Xu** – Thrust of Microelectronics of Function Hub, The Hong Kong University of Science and Technology (Guangzhou), Guangzhou, 511442, China; Email: renjingxu@hkust-gz.edu.cn

**Yingchun Cheng** – State Key Laboratory of Metastable Materials Science and Technology, Key Laboratory for Microstructural Material Physics of Hebei Province & School of Science, Yanshan University, Qinhuangdao 066004, China; Email: iamyccheng@ysu.edu.cn

**Xuetao Gan** – School of Integrated Circuits and Microelectronics & School of Physical Science and Technology, Northwestern Polytechnical University, Xi'an 710129, China; Email: xuetaogan@nwpu.edu.cn

**Wei Huang** – State Key Laboratory of Flexible Electronics & School of Flexible Electronics, Northwestern Polytechnical University, Xi'an 710129, China; Email: vc@nwpu.edu.cn

**Authors**

**Xiaolong Zhang** – State Key Laboratory of Flexible Electronics & School of Flexible Electronics, Northwestern Polytechnical University, Xi'an 710129, China;

**Honglei Chen** – Thrust of Microelectronics of Function Hub, The Hong Kong University of Science and Technology (Guangzhou), Guangzhou, 511442, China;

**Jiaming Wang** – State Key Laboratory of Flexible Electronics & School of Flexible Electronics, Northwestern Polytechnical University, Xi'an 710129, China;

**Fan Liu** – State Key Laboratory of Flexible Electronics & School of Flexible Electronics, Northwestern Polytechnical University, Xi'an 710129, China;

**Jiongtao Zhang** – State Key Laboratory of Flexible Electronics & School of Flexible Electronics, Northwestern Polytechnical University, Xi'an 710129, China;

**Junqiang Zhang** – State Key Laboratory of Flexible Electronics & School of Flexible Electronics, Northwestern Polytechnical University, Xi'an 710129, China;

**Yihan Yin** – School of Integrated Circuits and Microelectronics & School of Physical Science and Technology, Northwestern Polytechnical University, Xi'an 710129, China;

**Jinpeng Xu** – Institute of Physics, Henan Academy of Sciences, Zhengzhou 450046, China;

**Lei Ying** – School of Physics and Zhejiang Key Laboratory of Micro-nano Quantum Chips and Quantum Control, Zhejiang University, Hangzhou 310027, China

## Author Contributions

[#] X.L., X.Z., H.C., and J.W. contributed equally to this work. X.L. and R.X. conceived the concept and directed the collaboration and execution. X.L., R.X., Y.C., X.G. and W.H. supervised the research. X.Z., J.W., and F.L. fabricated the devices and performed the measurements. Jiongtao Zhang, Junqiang Zhang, Y.Y. and J.X. contributed to the device fabrication. X.L., J.W., and L.Y. analyzed the experimental data. H.C. and R.X. did the in-sensor simulation. X.L., H.C. and R.X. cowrote the manuscript with contributions from all the authors. All authors discussed the results and implications and commented on the manuscript at all stages.

## Notes

The authors declare no competing financial interest.

## ■ ACKNOWLEDGEMENTS

This work was supported by the National Natural Science Foundation of China (Nos. 12574459, 62405255 and 12375021), the Natural Science Basic Research Program of Shaanxi (No. 2025JC-YBMS-654), the Guangdong Basic and Applied Basic Research Foundation (No. 2023A1515110679), the Zhejiang Provincial Natural Science Foundation of China (No. LD25A050002), and the National Key Research and Development Program of China (No. 2022YFA1404203).

## ■ REFERENCES

Supporting information for

# Spatiotemporal programming via asymmetric dielectric engineering for nonvolatile 2D optoelectronics

Xiaoguang Luo[1, ☆, *], Xiaolong Zhang[1, ☆], Honglei Chen[2, ☆], Jiaming Wang[1, ☆], Fan Liu[1], Jiongtao Zhang[1], Junqiang Zhang[1], Yihan Yin[3], Jinpeng Xu[4], Lei Ying[5], Renjing Xu[2,*], Yingchun Cheng[6,*], Xuetao Gan[3,*], and Wei Huang[1,*]

[1] State Key Laboratory of Flexible Electronics & School of Flexible Electronics, Northwestern Polytechnical University, Xi'an 710129, China

[2] Thrust of Microelectronics of Function Hub, The Hong Kong University of Science and Technology (Guangzhou), Guangzhou, 511442, China

[3] School of Microelectronics & School of Physical Science and Technology, Northwestern Polytechnical University, Xi'an 710129, China

[4] Institute of Physics, Henan Academy of Sciences, Zhengzhou 450046, China

[5] School of Physics and Zhejiang Key Laboratory of Micro-nano Quantum Chips and Quantum Control, Zhejiang University, Hangzhou 310027, China

[6] State Key Laboratory of Metastable Materials Science and Technology, Key Laboratory for Microstructural Material Physics of Hebei Province & School of Science, Yanshan University, Qinhuangdao 066004, China

☆ These authors contributed equally to this work: Xiaoguang Luo, Jiaming Wang, Honglei Chen, Xiaolong Zhang

* Correspondence: Xiaoguang Luo (iamxgluo@nwpu.edu.cn), Renjing Xu (renjingxu@hkust-gz.edu.cn), Yingchun Cheng (iamyccheng@ysu.edu.cn), Xuetao Gan (xuetaogan@nwpu.edu.cn), or Wei Huang (vc@nwpu.edu.cn).

## Table of contents

## S1. Materials and methods

### Device fabrication

All 2D materials (Gr, hBN, $MoTe_2$, and $WSe_2$) were mechanically exfoliated from bulk crystals (HQ Graphene) using Scotch–tape (3M). The dual–floating–gate structures were patterned directly from large–scale Gr flakes on pre-cleaned 285-nm-thickness $SiO_2$/Si substrate (highly n–doped Si with resistivity < 0.01 Ω cm). Patterning was performed with a maskless ultraviolet lithography system (405 nm, ATS–07–UV Litho–ACA, TuoTuo Technology) using a bilayer photoresist stack of LOR 3A (Kayaku Advanced Materials) and S1805 (DuPont), followed by oxygen–plasma etching (PT500, LEBO Science) and a standard lift–off process. All other functional films, including the electrodes (~50 nm Au films), were assembled via a dry transfer technique (E1–T, MetaTest) with the assistance of polydimethylsiloxane (PDMS) stamp. Finally, the as–prepared devices were annealed at 200 °C for 1 h in a vacuum tube furnace (BTF–1200C–S, BEQ) to remove the resisted residues and reduce contact resistance.

### Device characterization

AFM and KPFM measurements were performed using an atomic force microscope (Dimension FastScan, Bruker). All electrical and optoelectrical characterizations were conducted at room temperature in a probe station under high vacuum (~$5\times10^{-6}$ mbar) using a semiconductor parameter analyzer (FS380 Pro, Platform Design Automation). A modulated 532 nm laser (spot diameter ~3 mm) provided the optical excitation for photocurrent measurements. Spatially resolved photocurrent mapping was carried out under ambient conditions using a home–built scanning system with a focused 532 nm laser spot (~3 μm in diameter).

### In–sensor ANN computing

The in-sensor ANN was simulated using the PyTorch framework, with a 128×128 pixel input array and a three-node output layer. The network was trained offline on the KTH Action Dataset. Input images were synthesized by sampling frames from each motion video. A quantization–aware training (QAT) strategy was employed to ensure the hardware compatibility, which constrains the synaptic weights to the nine discrete, normalized photoresponse levels achievable by the DFG devices. A straight–through estimator (STE) was used to approximate gradients during backpropagation, effectively incorporating device-specific physical constraints into the digital optimization process. The output layer utilized a softmax activation function, and the network was optimized using a cross-entropy loss function.

## S2. Fabrication of the dual-floating gate $MoTe_2$ devices

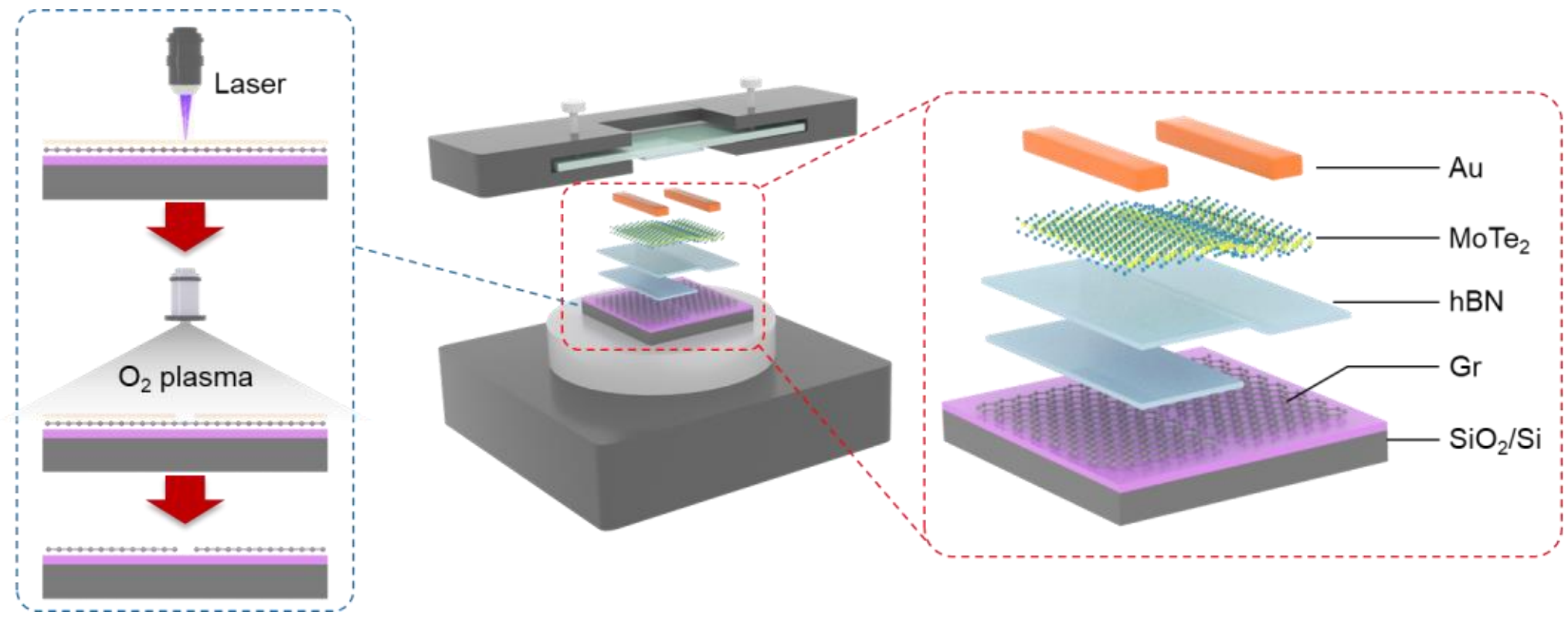

**Fig. S1. Device fabrication process**. Schematic diagrams of two key steps: (left) definition of the graphene dual-floating gate via maskless ultraviolet (405 nm) lithography and oxygen plasma etching; (Middle) assembly of the device structure using a PDMS-assisted dry transfer method. Right: exploded view schematic of the final layer stack.

The Gr dual–floating gates were fabricated as follows (with key processes illustrated in Fig. S1). First, a large–scale graphene flake was mechanically exfoliated onto a $SiO_2$/Si substrate using Scotch–tape. A bilayer photoresist stack was then applied: LOR 3A (Kayaku Advanced Materials) was spin–coated (EZ4–S spin coater, LEBO Science) at 3500 rpm for 50 s and baked at 160 °C for 5 min, followed by S1805 (DuPont) spin–coated at 4000 rpm for 60 s and a soft bake at 100 °C for 60 s. A predefined 1 μm gap pattern was exposed using a maskless ultraviolet lithography system. Development was performed for 11 s in positive photoresist developer (ZX-238). The final Gr dual–floating gates were obtained by oxygen–plasma etching and a standard lift–off process. For efficiency in this study, the Gr dual-floating gate in two devices are obtained directly during the mechanical exfoliation.

All 2D material flakes were mechanically exfoliated from commercially available bulk crystals with Scotch–tape. The heterostructures of devices were assembled via a dry transfer technique employing the polydimethylsiloxane (PDMS) stamp, as illustrated in Fig. S1. Onto the prefabricated Gr dual–floating gates, we sequentially transferred two hBN flakes, one $MoTe_2$ flake, and two Au electrodes. Finally, the as–prepared devices were annealed at 200 °C for 1 h in a vacuum tube furnace to remove polymeric residues and reduce contact resistance.

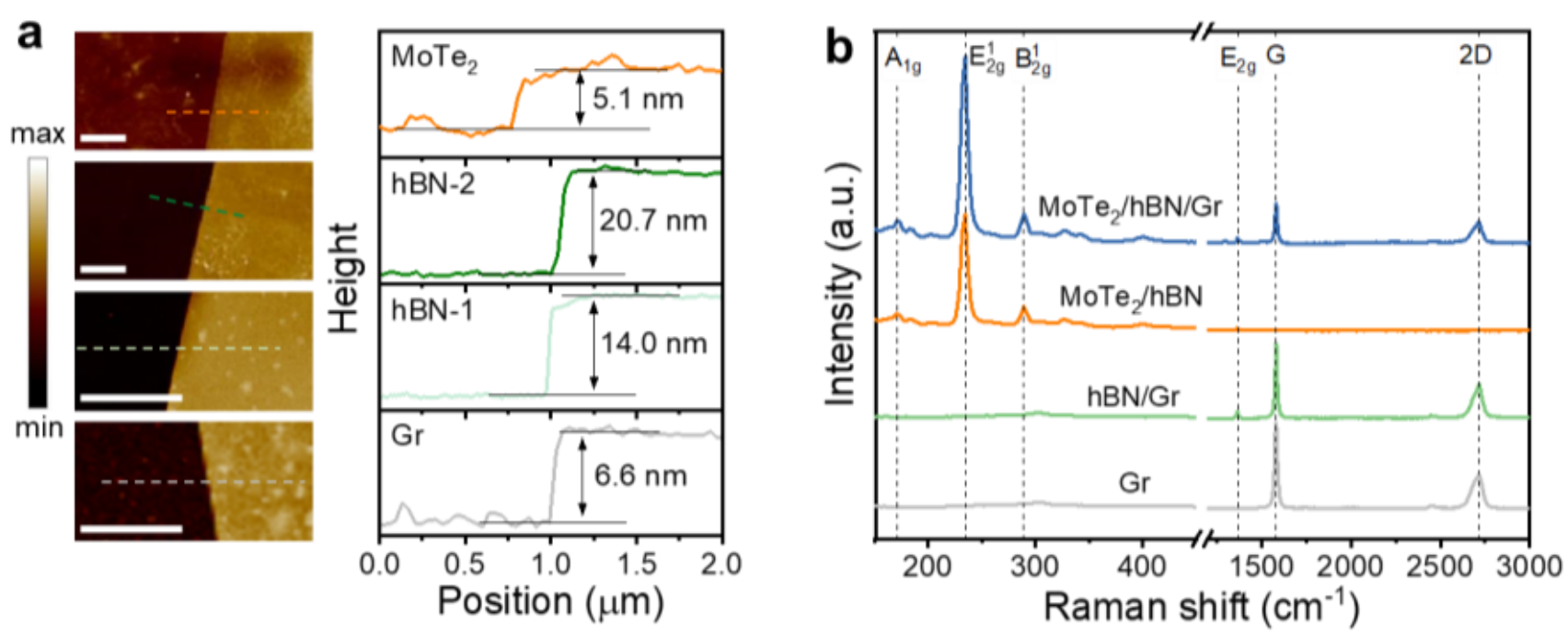

**Fig. S2. AFM and Raman scattering characterizations of the device shown in Fig. 1b. a)** AFM mapping (left) and height profile (right) related to $MoTe_2$ (5.1 nm), top hBN-2 (20.7 nm), bottom hBN-1 (14.0 nm), and graphene (6.6 nm) nanosheets, scale bar: 1 μm. **b)** Raman scattering spectra of 2D materials or heterostructures in the device, where peaks at 171, 234.5, and 289.2 $cm^{-1}$ account for the out-of-plane $A_{1g}$, in-plane $E_{2g}^{1}$ modes, and out-of-plane (active in few-layer) $B_{2g}^{1}$ vibration modes of $MoTe_2$, peaks at 1368 $cm^{-1}$ correspond to the $E_{2g}$ mode of hBN, and peaks at around 1580.5 and 2714 $cm^{-1}$ are ascribed to the G band and 2D band of graphene, respectively.

Raman spectra of $MoTe_2$, hBN, and Gr were measured by a confocal Raman microscope (Alpha300R, WITec) under ambient conditions. The samples were excited with a 532 nm laser (spot diameter ~400 nm, laser power ~1 mW, spectral resolution ~0.02 $cm^{-1}$). As shown in Fig. S2b, the sharp characteristic peaks observed in the Raman spectra confirm the high crystallinity of the 2D materials. AFM was used to measure the flake thickness (Fig. S2a) of the devices. KPFM was employed to map the surface potential distribution (Fig. 1e) of devices previously programmed with gate–voltage dipulses.

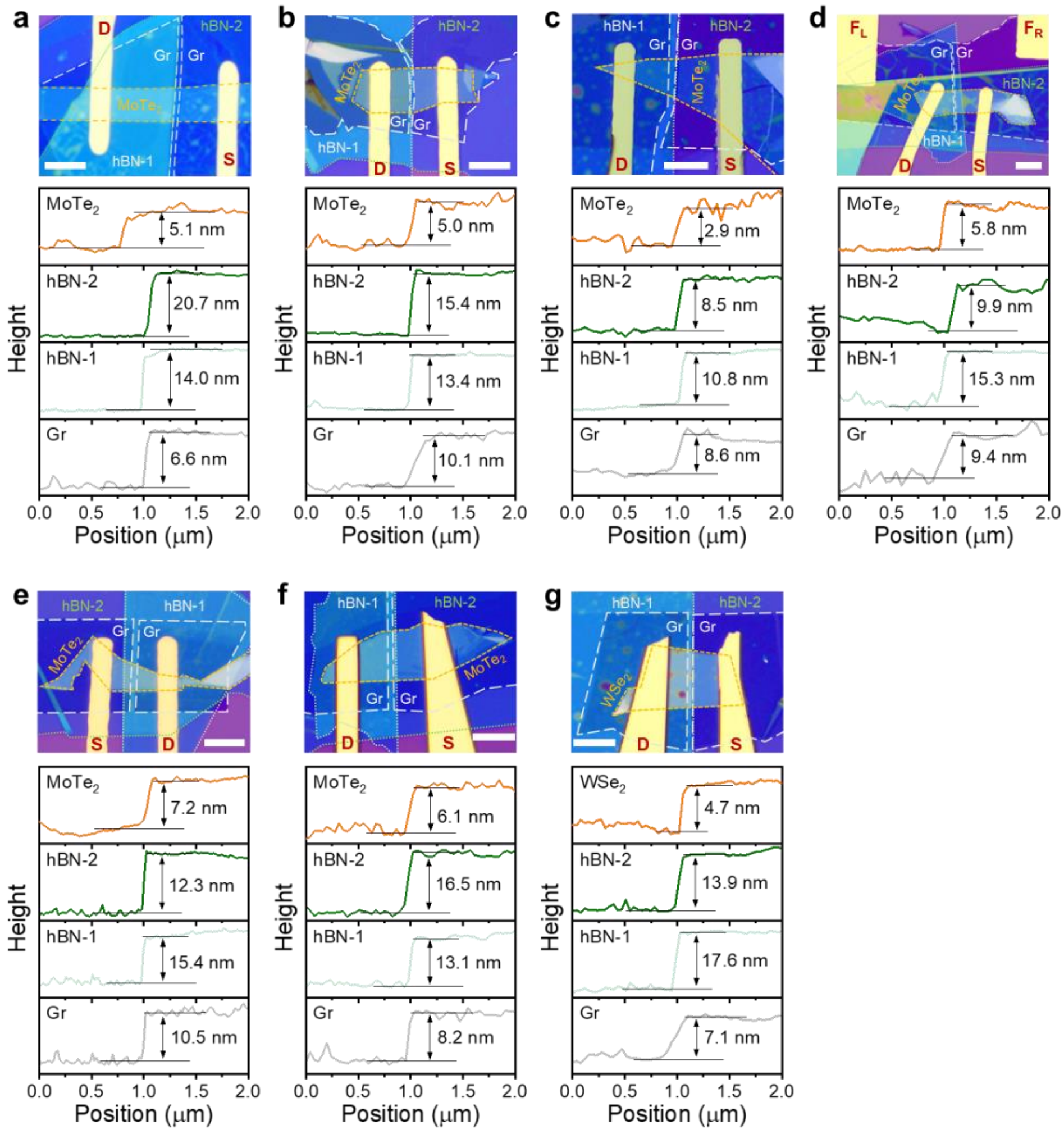

**Fig. S3. Fabricated device library.** Optical images (upper row) and corresponding AFM height profiles (lower row) of the key devices used in this study. Each device was fabricated for targeted experiments: **a)** Electrical and optoelectronic characterizations (Figs. 1c-1e, 2a, 4a, S2, S6, S7, S12, and S13). **b)** KPFM characterizations (Figs. 1f and S4). **c)** Verification of homojunction reconfiguration (Fig. 2d). **d)** Thickness-dependent FN tunneling in a single device (Fig. S8) and long-term retention test after 7 months (Fig. S10). **e)** Long-term retention test (Fig. 3a) and rectifier applications (Fig. 3b). **f)** Programmable photoresponse (Figs. 4b and 4c). **g)** Universality demonstration with ambipolar $WSe_2$. Scale bar: 10 μm.

## S3. KPFM characterizations of the device

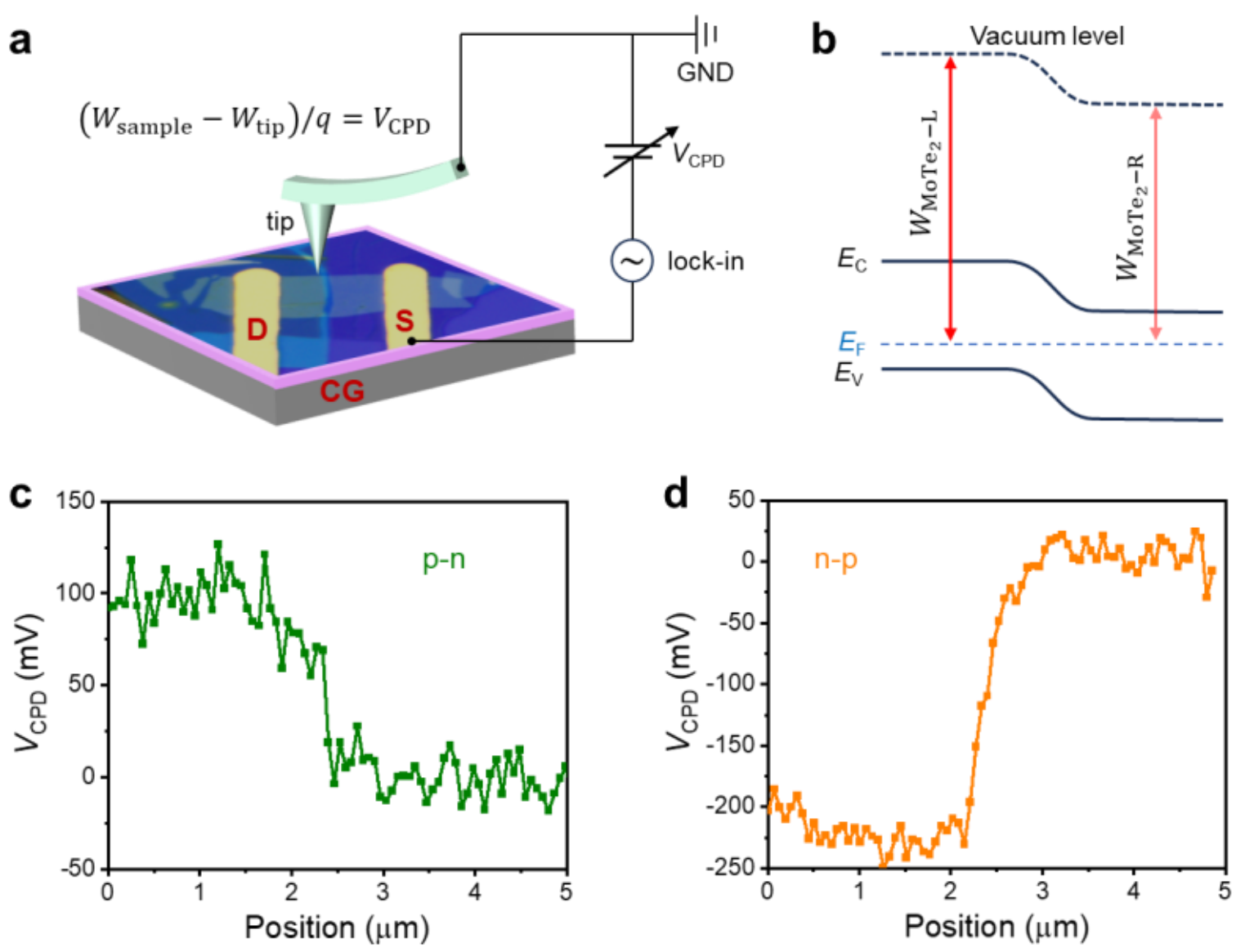

**Fig. S4. KPFM characterizations of a MoTe₂/hBN/Gr DFG device. a)** Schematic of the KPFM measurement setup, where $W_{sample/tip}$ denote the work function of the sample/tip, and $V_{CPD}$ is the contact potential difference in volts. In the measured devices, the left semi-channel is positioned over the dielectric composed of two hBN flakes. **b)** Schematic band diagram illustrating the work functions in the $MoTe_2$ p–n homojunction. $V_{CPD}$ profiles across the homojunction along the dashed guidelines of Fig. 1e in the **c)** p–n configuration ($V_1 = 70$ V, $V_2 = -18$ V) and **d)** n–p configuration ($V_1 = -70$ V, $V_2 = 20$ V).

In the KPFM measurement setup (Fig. S4a), a voltage is applied to the sample. The contact potential difference ($V_{CPD}$) is related to the work functions by:

$$(W_{sample} - W_{tip})/q = V_{CPD}$$

where the work function is defined as the energy difference between Fermi level and vacuum level. In the p–n configuration (Fig. S4b), the work function on the left side of the $MoTe_2$ homojunction is larger than on the right ($W_{MoTe_2-L} > W_{MoTe_2-R}$). From the relationship $(W_{MoTe_2-L} - W_{MoTe_2-R})/q = V_{CPD-L} - V_{CPD-R}$, it follows that $V_{CPD-L} > V_{CPD-R}$ for p–n configuration and $V_{CPD-L} < V_{CPD-R}$ for n–p configuration, in great agreement with the experimental data (Figs. S4c and S4d).

## S4. Duration of gate voltage pulses for the DFG $MoTe_2$ devices

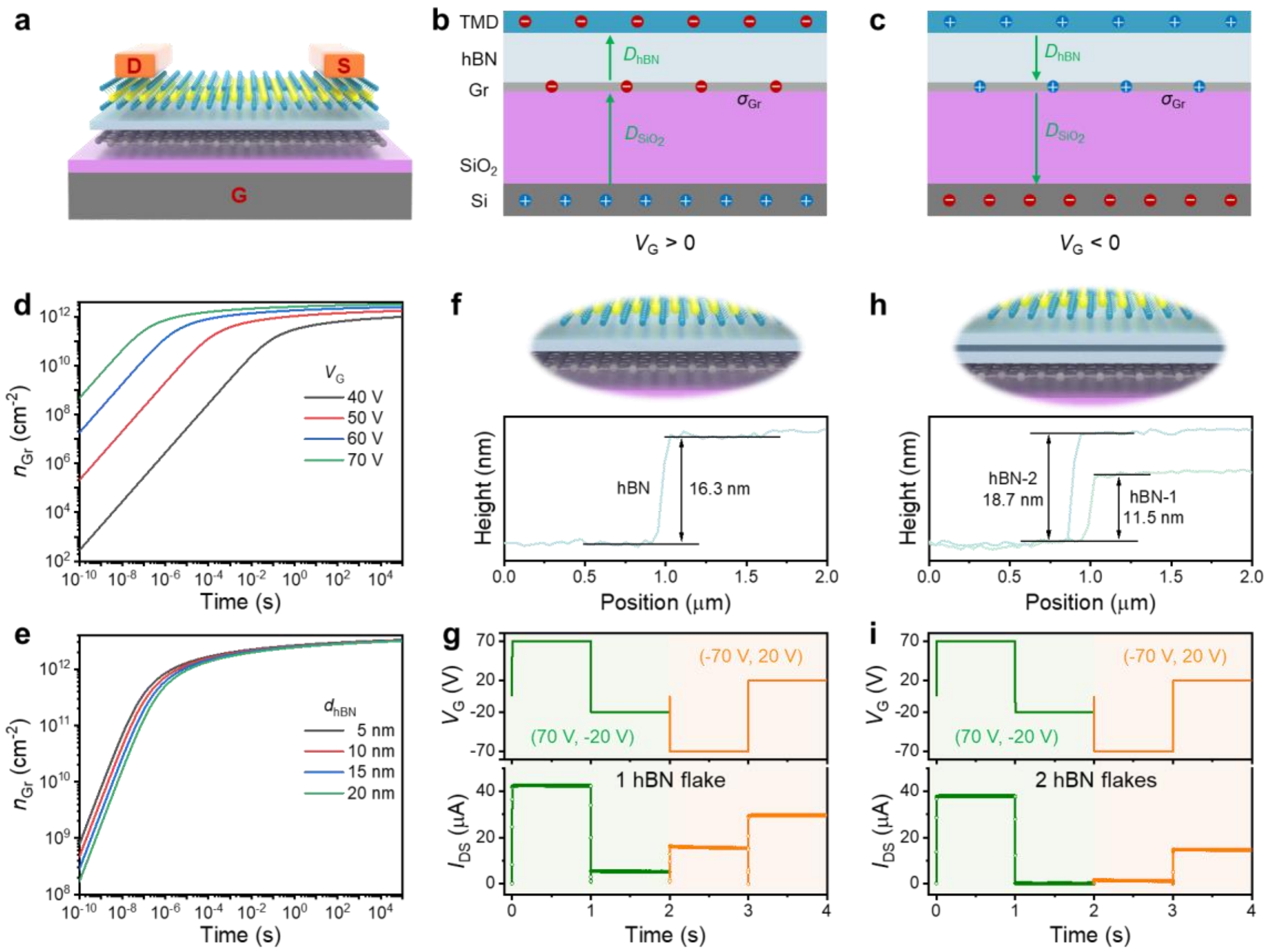

**Fig. S5. Duration of gate voltage pulses for the $MoTe_2$/hBN/Gr single-floating gate device**. **a)** Schematic illustrations of the single-floating-gate transistor. **b-c)** The corresponding charge distribution when $V_{\mathrm{G}} > 0$ and $V_{\mathrm{G}} < 0$, respectively, where $D$ denotes the displacement field and $\sigma$ the charge density. **d)** Calculated electron densities trapped in the Gr floating gate versus the duration time of gate voltage pulse for a 10 nm thick hBN dielectric. **e)** Calculated electron densities trapped in the Gr floating gate with the duration time of gate voltage pulse for different hBN thickness, where the gate voltage is 70 V. **f)** Device with the dielectric consisting of a 16.3 nm thick hBN flake. **g)** The transient current of the device (lower) with single hBN flake at $V_{\mathrm{DS}} = 1$ V, under the corresponding gate voltage pulses (upper). **h)** Device with the dielectric comprising double hBN flakes (11.5 and 18.7 nm thick). **i)** The transient current (lower) and corresponding gate voltage pulses (upper) for the double-hBN-flake device when $V_{\mathrm{DS}} = 1$ V.

Focus on the $MoTe_2$/hBN/Gr single-floating gate transistor (Fig. S5a), the Gr floating gate will trap electrons and holes under the positive and negative gate voltages $V_{\mathrm{G}}$, as shown in Figs. S5b-S5c. In the $SiO_2$ and hBN dielectrics, $D_{\mathrm{SiO_2}} = \varepsilon_{\mathrm{SiO_2}} E_{\mathrm{SiO_2}}$ and $D_{\mathrm{hBN}} = \varepsilon_{\mathrm{hBN}} E_{\mathrm{hBN}}$, where $D$ denotes the displacement field, $E$ the electric field, and $\varepsilon$ the dielectric constant. According to Gauss theorem, we have:

$$\begin{cases} D_{\mathrm{hBN}} - D_{\mathrm{SiO_2}} = -\sigma_{\mathrm{Gr}} \\ d_{\mathrm{SiO_2}} E_{\mathrm{SiO_2}} + d_{\mathrm{hBN}} E_{\mathrm{hBN}} = V_{\mathrm{G}} \end{cases}$$

Here $\sigma_{\mathrm{Gr}}$ is treated as the positive value for convenience. The electric field in hBN layer can be

solved as [1]:

$$E_{\mathrm{hBN}} = \frac{\varepsilon_{\mathrm{SiO_2}} V_{\mathrm{G}} - d_{\mathrm{SiO_2}} \sigma_{\mathrm{Gr}}}{d_{\mathrm{SiO_2}} \varepsilon_{\mathrm{hBN}} + \varepsilon_{\mathrm{SiO_2}} d_{\mathrm{hBN}}} = \frac{V_{\mathrm{G}}}{d_{\mathrm{SiO_2}} \frac{\varepsilon_{\mathrm{hBN}}}{\varepsilon_{\mathrm{SiO_2}}} + d_{\mathrm{hBN}}} - \frac{\sigma_{\mathrm{Gr}}}{\varepsilon_{\mathrm{hBN}} + \varepsilon_{\mathrm{SiO_2}} \frac{d_{\mathrm{hBN}}}{d_{\mathrm{SiO_2}}}}$$

where $d$ is the layer thickness. The first term denotes the initial electric field in hBN layer $E_0 = V_{\mathrm{G}}/\left(d_{\mathrm{SiO_2}} \varepsilon_{\mathrm{hBN}}/\varepsilon_{\mathrm{SiO_2}} + d_{\mathrm{hBN}}\right)$ that produced by gate voltage $V_{\mathrm{G}}$, and the second term indicates the induced electric field by the trapped charges. Noticed that both $E_{\mathrm{hBN}}$ and $\sigma_{\mathrm{Gr}}$ are of time-dependent, and therefore the variation is $dE_{\mathrm{hBN}} = -d\sigma_{\mathrm{Gr}}/B_0$ with $B_0 = \varepsilon_{\mathrm{hBN}} + \varepsilon_{\mathrm{SiO_2}} d_{\mathrm{hBN}}/d_{\mathrm{SiO_2}}$.

Based on the Fowler-Nordheim tunneling theory, the tunneling current through hBN is [2]

$$J = C_0 E_{\mathrm{hBN}}^2 e^{-D_0/E_{\mathrm{hBN}}} = \frac{d\sigma_{\mathrm{Gr}}}{dt}$$

where $C_0 = \frac{q^3 m_{\mathrm{emitter}}}{8\pi h \Phi_{\mathrm{B}} m_{\mathrm{hBN}}}$ and $D_0 = \frac{8\pi\sqrt{2 m_{\mathrm{hBN}} \Phi_{\mathrm{B}}^3}}{3hq}$, with the electron charge $q$, Planck constant $h$, $m$ carrier effective mass, and barrier height $\Phi_{\mathrm{B}}$ between electron emitter material and hBN. $MoTe_2$ and Gr are the electron emitter materials for $V_{\mathrm{G}} > 0$ and $V_{\mathrm{G}} < 0$, respectively. The solution of the above equation is

$$E(t) = \frac{D_0}{\ln\left(e^{\frac{D_0}{E_0}} + \frac{C_0 D_0}{B_0} t\right)}$$

and the trapped charge density is

$$n_{\mathrm{Gr}} = \frac{\sigma_{\mathrm{Gr}}}{q} = \frac{B_0[E_0 - E(t)]}{q}$$

The calculation for $V_{\mathrm{G}} > 0$ has been conducted with the given parameters of $m_{\mathrm{hBN}} = 0.26 m_0$ [3], $m_{\mathrm{MoTe_2}} = 0.64 m_0$ [4], $\Phi_{\mathrm{B}} \approx 1.5$ eV [5,6], $\varepsilon_{\mathrm{SiO_2}} = 3.9\varepsilon_0$, $\varepsilon_{\mathrm{hBN}} = 2.5\varepsilon_0$, and the vacuum dielectric constant $\varepsilon_0$, as shown in Fig. S5d-S5e. The results indicate that the trapping of charge in Gr floating gate saturates more rapidly at higher gate voltages and, to a lesser extent, with a thinner hBN dielectric layer. The saturation can be further accelerated by the lower $\Phi_{\mathrm{B}}$ and $\varepsilon_{\mathrm{hBN}}$ (not shown here), promising that experimental operation in the nanosecond regime is achievable for such hBN/Gr single-floating gate memory devices [7,8].

In our devices, single and double hBN flakes act as the dielectric layers above two split Gr floating gates, respectively. To ensure stable device operation, a fixed gate-voltage pulse duration, $\Delta t$, must be established. Here, we examine two distinct $MoTe_2$/hBN/Gr single-floating gate transistors: one with a 16.3 nm thick hBN layer (Fig. S5f) and the other with a dielectric stack comprising two hBN flakes (11.5 and 18.7 nm thick; Figs. S5h). Using a pulse duration of $\Delta t = 1$ s, we measured the transient current in both devices under dipulse of gate voltages, including the set of (70, -20) V and (-70, 20) V. Results show that transient current saturates rapidly, as measured by our source meter, confirming that $\Delta t = 1$ s is fully adequate for charge trapping saturation in our $MoTe_2$/hBN/Gr dual-floating gate devices.

## S5. Electrical characterizations of the devices

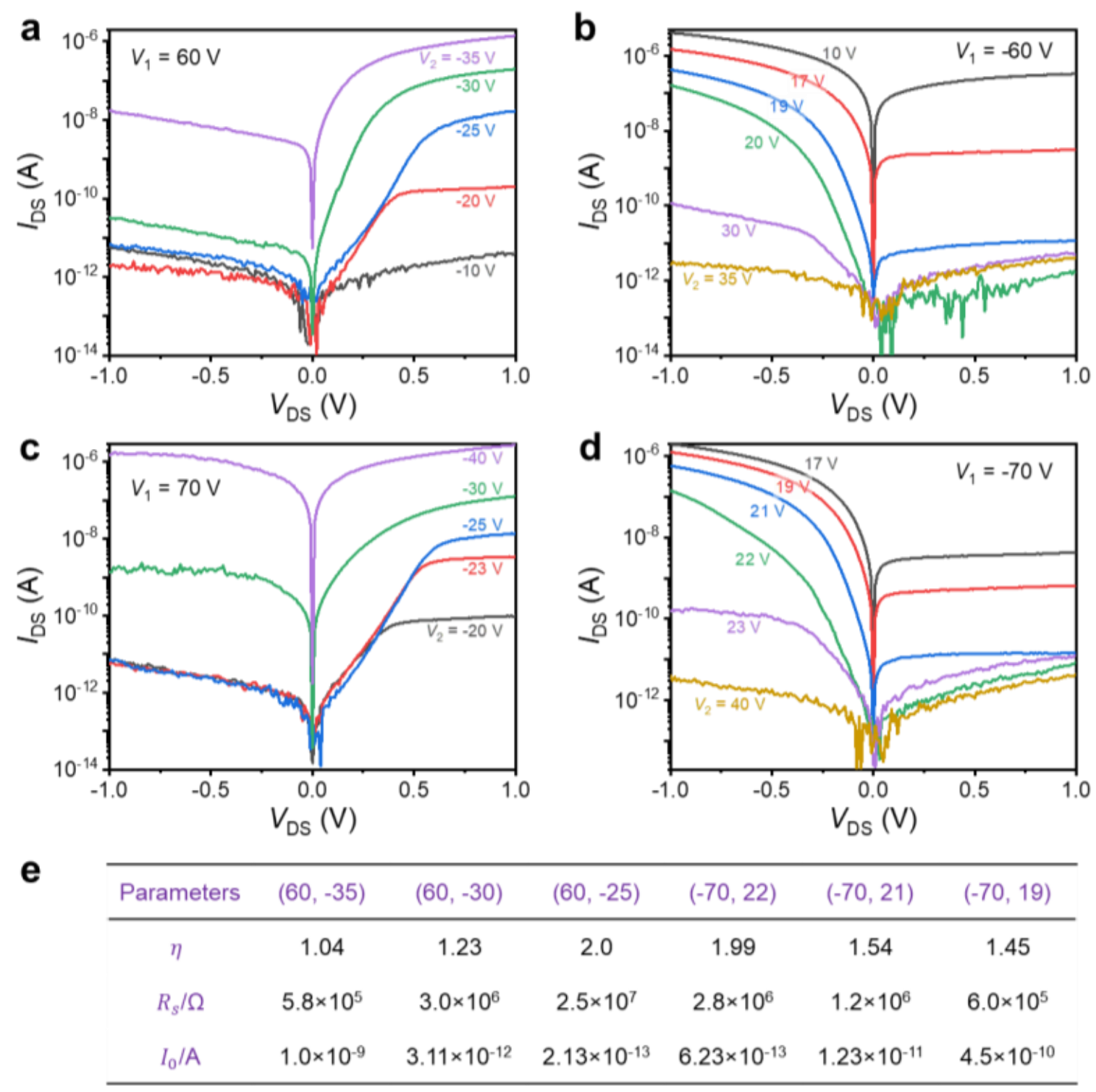

| Parameters | (60, -35) | (60, -30) | (60, -25) | (-70, 22) | (-70, 21) | (-70, 19) |
|---|---|---|---|---|---|---|
| $\eta$ | 1.04 | 1.23 | 2.0 | 1.99 | 1.54 | 1.45 |
| $R_s/\Omega$ | $5.8\times10^{5}$ | $3.0\times10^{6}$ | $2.5\times10^{7}$ | $2.8\times10^{6}$ | $1.2\times10^{6}$ | $6.0\times10^{5}$ |
| $I_0$/A | $1.0\times10^{-9}$ | $3.11\times10^{-12}$ | $2.13\times10^{-13}$ | $6.23\times10^{-13}$ | $1.23\times10^{-11}$ | $4.5\times10^{-10}$ |

**Fig. S6. Output characteristics of the $MoTe_2$/hBN/Gr DFG device**. **a-d)** Output curves for an initial gate-voltage pulse of 60, -60, 70, and -70 V, respectively, with the second voltage pulse being labeled in each figure. **e)** Fitting results by the modified Shockley equation with Lambert function for selected voltage sets of dipulse input.

Figs. S6a-S6d present the output characteristic curves for the initial voltage pulse of $|V_1| = 60$ and 70 V, respectively. The rectification behavior is observed for all dipulse inputs, which is also modulated by the specific voltage sets of dipulse inputs. The rectification direction reverses between cases with positive and negative voltages of the first pulse, indicating a reconfiguration of the junction type within the device. Accounting for the series resistance ($R_s$) of the $MoTe_2$ channel, the output characteristics can be expressed by a Lambert $W$-function based modified Shockley

equation:

$$I_{\mathrm{DS}} = \frac{\eta V_{\mathrm{T}}}{R_s} W \left[\frac{I_0 R_s}{\eta V_{\mathrm{T}}} e^{(V_{\mathrm{DS}}+I_0 R_s)/\eta V_{\mathrm{T}}}\right] - I_0$$

where $V_{\mathrm{T}} = k_{\mathrm{B}} T/q$ is the thermal voltage at temperature $T$, $k_{\mathrm{B}}$ is the Boltzmann constant, $I_0$ is the reverse saturation current, and $\eta$ is the ideality factor. The experimental data for voltage sets of (60,-35), (60,-30), (60,-25), (-70,22), (-70,21), and (-70,19) V were fitted with this model, yielding ideality factors between 1 and 2 (Fig. S6e).

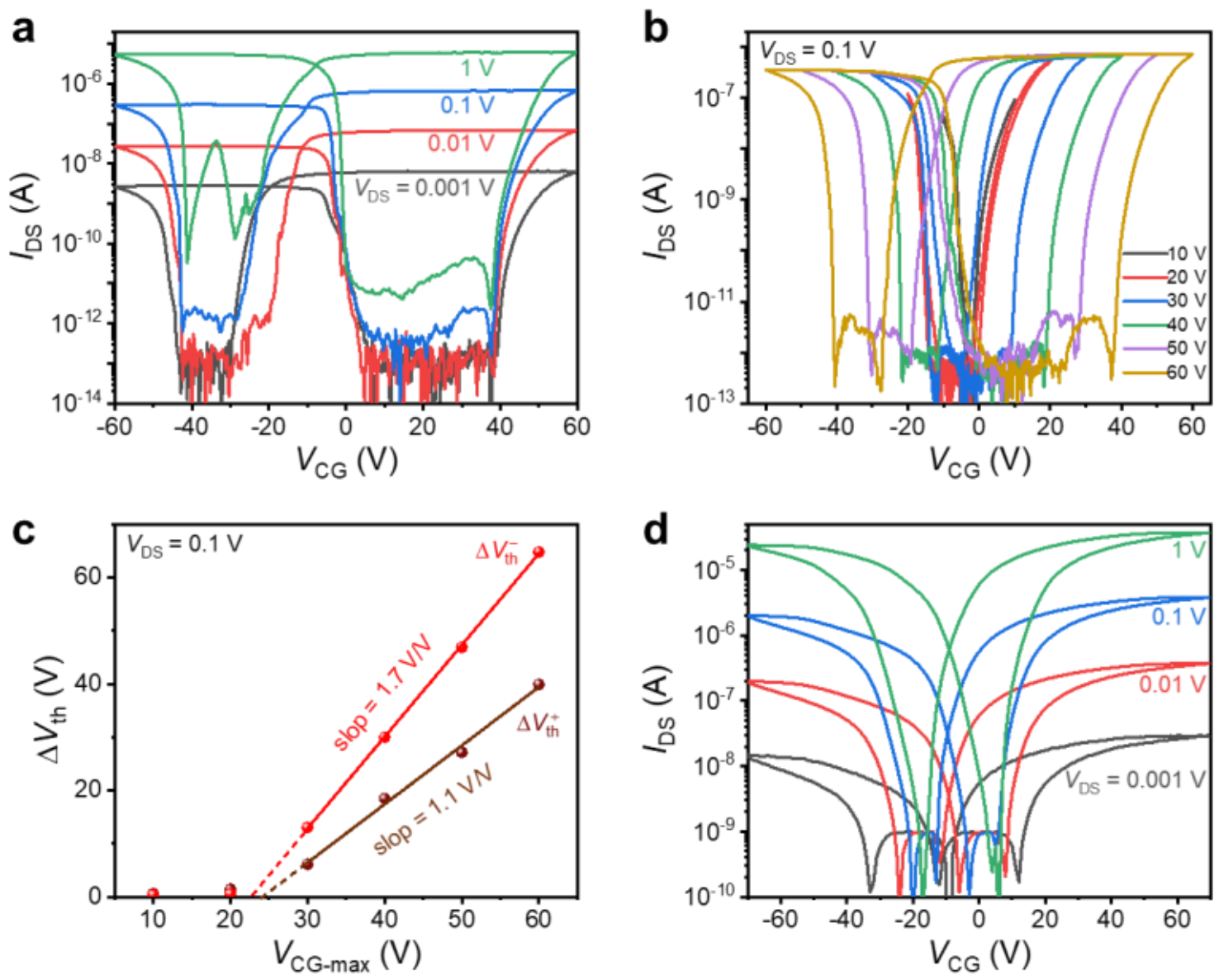

**Fig. S7. Transfer characteristics of the $MoTe_2$/hBN/Gr DFG device**. **a)** Dual–sweep transfer curves at the drain–source voltage of 0.001, 0.01, 0.1, and 1 V, respectively. **b)** Transfer curves at the fixed $V_{\mathrm{DS}} = 0.1$ V for different control–gate ($V_{\mathrm{CG}}$) sweep range. **c)** Corresponding memory windows in gate voltage for n–branch $\Delta V_{\mathrm{th}}^-$ and p–branch $\Delta V_{\mathrm{th}}^+$. **d)** Dual–sweep transfer curves of the single-floating gate device (Fig. S5f) at different drain–source voltages.

Fig. S7a presents the dual–sweep transfer curves of a $MoTe_2$/hBN/Gr DFG device at different drain-source voltages, revealing consistent memory windows. The memory windows in the n–branch ($\Delta V_{\mathrm{th}}^-$) are always larger than that in the p–branch ($\Delta V_{\mathrm{th}}^+$). This asymmetry between p– and n–branches persists when modulating the gate-voltage sweep range (Fig. S7b). The extracted values of $\Delta V_{\mathrm{th}}^+$ and $\Delta V_{\mathrm{th}}^-$ are plotted against the gate-voltage sweep range in Fig. S7c. Beyond a threshold, both $\Delta V_{\mathrm{th}}^+$ and $\Delta V_{\mathrm{th}}^-$ increase linearly with the gate-voltage sweep range, with slopes of 1.1 V/V for the p–branch and 1.7 V/V for the n–branch, respectively. For comparison, we also measured the dual–sweep transfer curves of a $MoTe_2$/hBN/Gr single-floating gate device (shown in Fig. S5f) at different drain-source voltage. In contrast, no pronounced asymmetry in memory windows is observed.

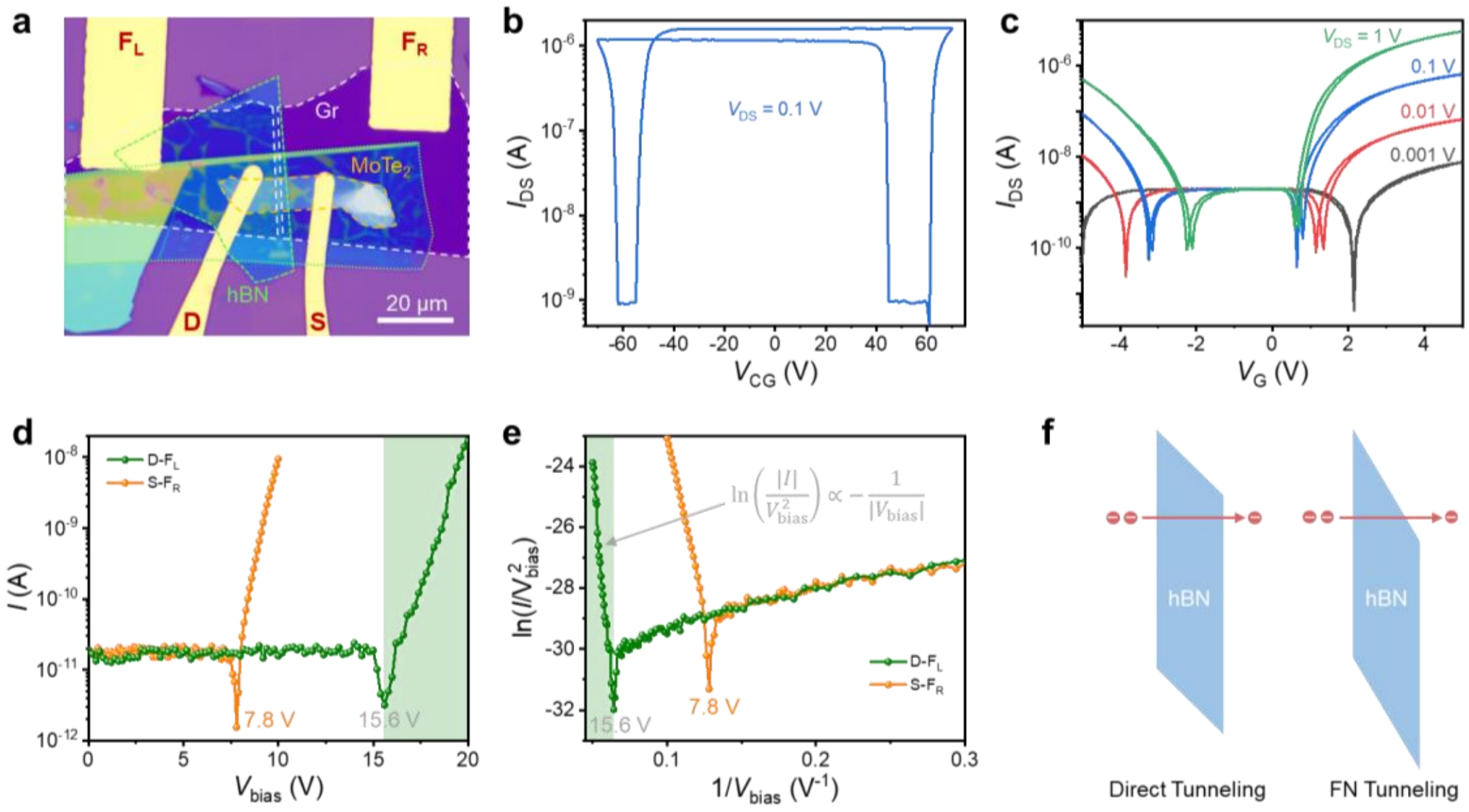

**Fig. S8. Floating-gate operation and FN tunneling in the $MoTe_2$/hBN/Gr DFG device**. **a)** Optical image of the device. Subscripts "L" and "R" denote the left and right floating gates, respectively. **b)** Transfer curves with the gate voltage applied to the Si substrate. **c)** Transfer curves with the gate voltage applied simultaneously to both Gr floating–gate layers. **d)** Tunneling current as a function of bias voltage between D/S and $F_{L/R}$ electrodes. **e)** Corresponding FN plots of the tunneling current. The FN tunneling regime for the dielectric of two hBN flakes is highlighted in green box. **f)** Schematic band diagram of direct and FN electron tunneling through the hBN dielectric.

The device in Fig. S8a was used to confirm the function of Gr layers in the floating–gate effect. Compared to applying the gate voltage to the control gate, the memory window vanished dramatically when the gate voltage is applied simultaneously to both Gr floating gates (Figs. S8b and S8c), directly verifying that the charge trapping occurs in Gr layers. We probed the tunneling behavior for each Gr floating gate by measuring the current with a bias voltage applied between D/S electrode and the respective underlying Gr floating gates, thereby investigating tunneling through either single or double hBN flakes in a single device. The obtained current–voltage characteristic (Fig. S8d) reveals two distinct tunneling regimes. These correspond to direct tunneling and FN tunneling (Fig. S8f), which are described by the following relationships, respectively [9]:

$$\ln\left(\frac{|I|}{V_{\text{bias}}^2}\right) \propto \ln\left(\frac{1}{|V_{\text{bias}}|}\right) - \frac{2d_{\text{hBN}}\sqrt{2m\Phi_{\text{B}}}}{\hbar}$$

and

$$\ln\left(\frac{|I|}{V_{\text{bias}}^2}\right) \propto -\frac{1}{|V_{\text{bias}}|}\frac{4d_{\text{hBN}}\sqrt{2m\Phi_{\text{B}}^3}}{3\hbar q}$$

Fitting with the experimental data confirms that direct tunneling dominates at low bias voltage,

while FN tunneling prevails at high bias. For the double–hBN–flake case (Fig. S8e), the transition between these regimes occurs at 15.6 V, as evidenced by the switch in functional dependence of the FN plot. The transition voltage is higher for the double–hBN–flake case than for the single–hBN–flake case (7.8 V). Moreover, the transition voltage increases with the total hBN thickness, a trend that holds for both single– and double–hBN–flake dielectrics, as shown in Fig. S9.

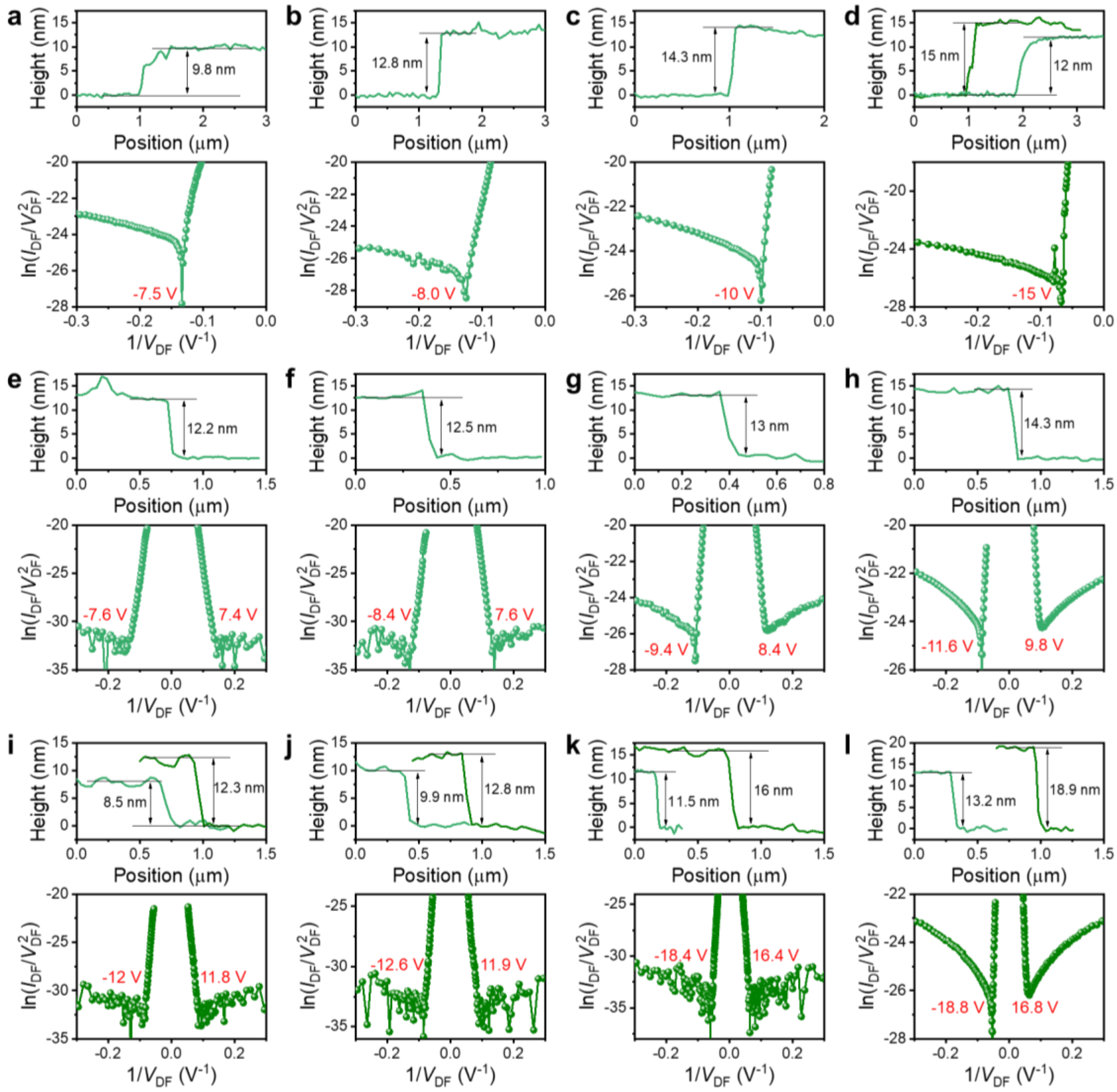

**Fig. S9. FN tunneling in $MoTe_2$/hBN/Gr heterostructures**. Upper panel: AFM height profile of the hBN dielectric, where single or double steps correspond to devices with single or double hBN flakes, respectively; lower panel: corresponding FN plot of the tunneling current. The transition voltages between direct and FN tunneling regimes are highlighted in red.

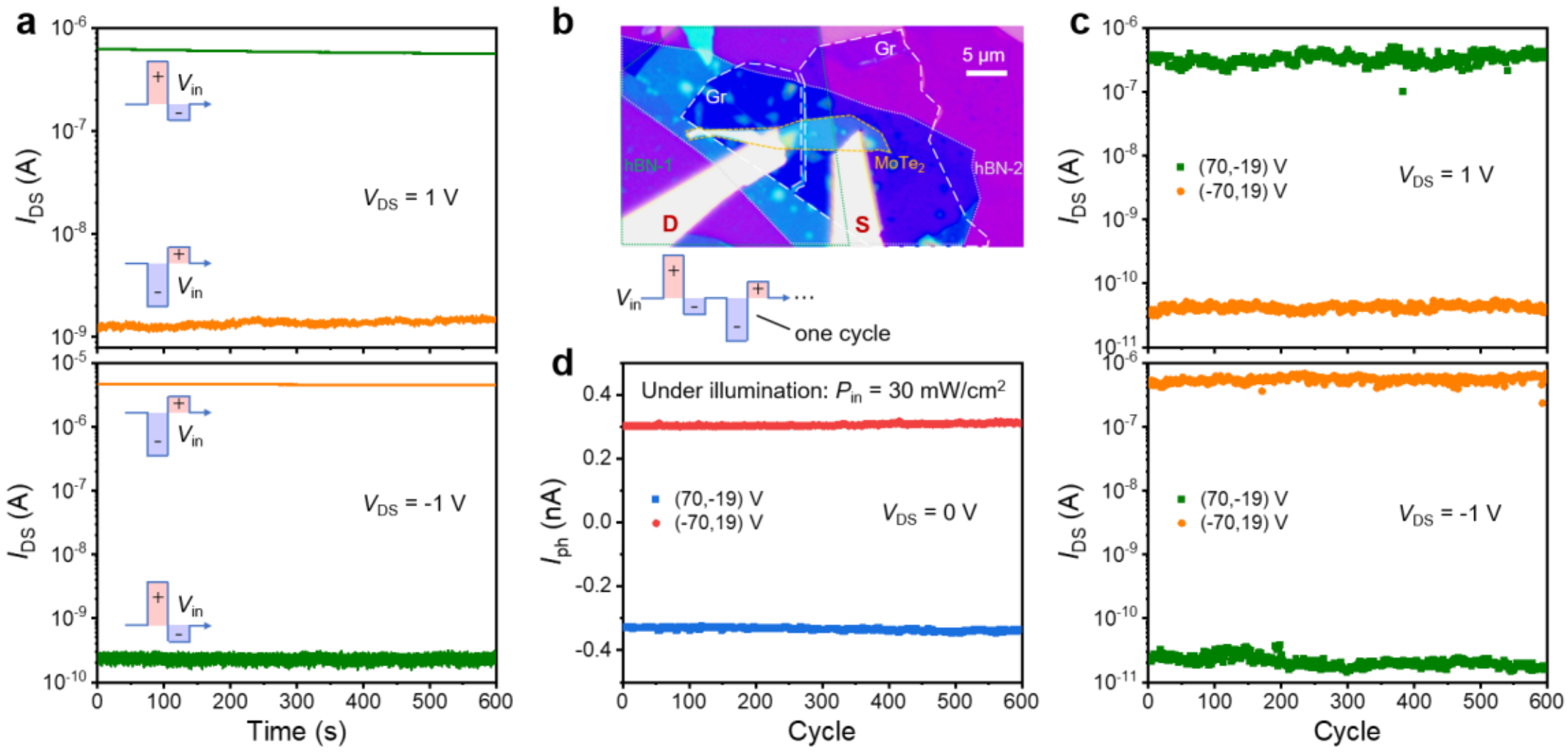

**Fig. S10. Reliability characterization of the DFG device. a)** Long-term retention of a device fabricated seven months ago. Transient current measured at $V_{DS} = 1$ V (upper) and $V_{DS} = -1$ V (lower) for the device programmed into p–n (70 V, -30 V) and n–p (-70 V, 25 V) states. **b)** Optical image of a device used for endurance testing (upper) and the input voltage profile for one measurement cycle (lower). **c)** Endurance characteristics of current for p–n (70 V, -19 V) and n–p (-70 V, 19 V) states, measured at $V_{DS} = 1$ V (upper) and $V_{DS} = -1$ V (lower). Each cycle consists of six voltage steps (70 V, -19 V, 0 V, -70 V, 19 V, 0 V) and each step lasts 1 s, with two currents sampled during the two 0-V steps. **d)** Endurance characteristics of photocurrent for p–n (70 V, -19 V) and n–p (-70 V, 19 V) states at $V_{DS} = 0$ V, measured under continuous 532 nm illumination at the laser power density of ~30 mW/cm$^2$.

## S6. Reconfigurable photoresponse of the devices

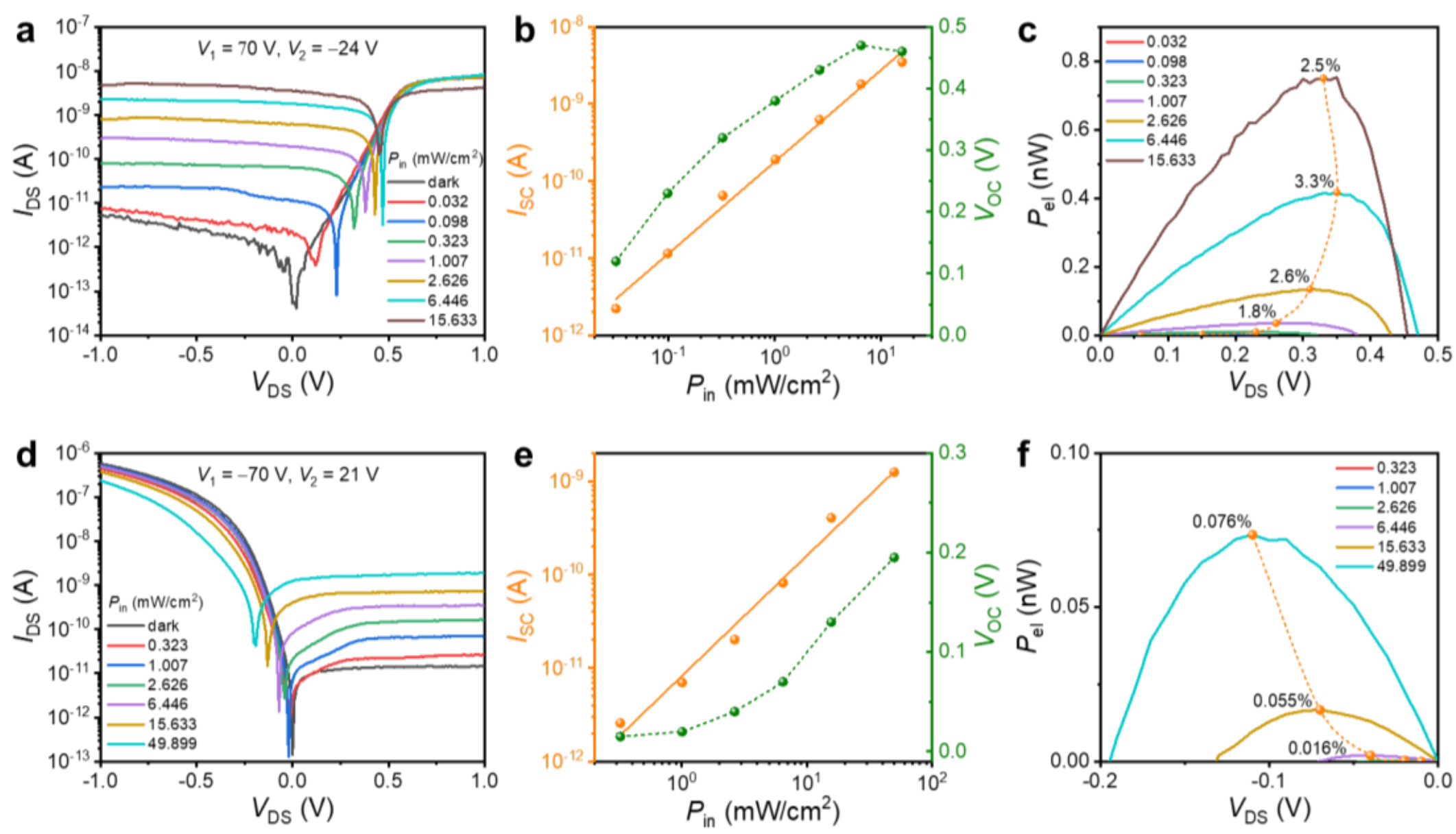

**Fig. S11. Reconfigurable photovoltaic effect**. **a-c)** Photovoltaic performance of the device under laser illumination (0–15.633 mW/cm$^2$) for the p–n state programmed by a (70, -24) V dipulse: output curves **a)**, extracted $I_{SC}$ and $V_{OC}$ **b)**, and calculated electrical power **c)**. **d-f)** Corresponding performance under laser illumination (0–49.899 mW/cm$^2$) for the n–p state programmed by a (-70, 21) V dipulse.

As programmed into p–n or n–p configuration, the $MoTe_2$/hBN/Gr DFG device should perform the photovoltaic effect due to the strong built–in electric field of the $MoTe_2$ homojunction. The photovoltaic characteristics for the p–n configuration, programmed by (70, –24) V dipulse, are presented in Figs. S11a-S11c. Under uniform 532 nm laser illumination (spot diameter ~3 mm), the output curves shift in the positive direction from zero with the well-defined short-circuit current ($I_{SC}$) and open-circuit voltage ($V_{OC}$), characteristic of an obvious photovoltaic response. Both $I_{SC}$ and $V_{OC}$ (voltage at the deep in Fig. S11a) increases with the laser power density ($P_{in}$), of which $V_{OC}$ exhibits saturation at strong illumination (Fig. S11b). The measured maximum $I_{SC}$ and $V_{OC}$ are 3.5 nA and 0.47 V, respectively. The electrical power,

$$P_{el} = I_{DS} \cdot V_{DS}$$

was calculated from the output curves, as shown in Fig. S11c. For all measured $P_{in}$, $P_{el}$ increases initially to a maximum value ($P_{el-max}$) before decreasing to zero at $V_{OC}$. $P_{el-max}$ reaches 0.75 nW at $P_{in} = 15.633$ mW/cm$^2$. Generally, the maximum electrical power can be further used to calculate the power conversion efficiency (PCE, $\eta_{PCE}$) of the device, i.e.,

$$\eta_{\mathrm{PCE}} = \frac{P_{\mathrm{el-max}}}{P_{\mathrm{in}} \cdot A}$$

where $A = 194.7$ μ$m^2$ is the $MoTe_2$ channel area. The PCE initially increases from 0.16% to 3.3%, and then decreases to 2.5%, as $P_{\mathrm{in}}$ increases to 15.633 mW/$cm^2$.

For the n–p configuration programmed by (–70, 21) V dipulse, a clear photovoltaic effect is also observed. Different from p–n configuration, the output curves shift in the negative direction (Fig. S11d), consistent with the band bending inversion upon reconfiguration. The photovoltaic parameters show a similar trend with increasing $P_{\mathrm{in}}$ (up to 49.899 mW/$cm^2$), yielding the measured maximum values of $I_{\mathrm{SC}} = 1.2$ nA, $V_{\mathrm{OC}} = 0.2$ V, $P_{\mathrm{el-max}} = 0.074$ nW, and $\eta_{\mathrm{PCE}} = 0.076\%$ (Figs. S11e and S11f).

To specify the photodetection performance of the $MoTe_2$/hBN/Gr DFG device, we conducted the transient photocurrent measurement through one–time test. The experimental setup is shown in Fig. S12a, where a stepped neutral density filter is used in the power–dependent characterization, and the 532 nm laser is modulated by an acousto–optic modulator at 1 Hz with a 40% duty cycle. Fig. S12b displays the transient photocurrent for the p–n configuration programmed with a (70, –23) V dipulse at $V_{\mathrm{DS}} = 0$ V, when laser power density increases from 0.057 to 74.791 mW/$cm^2$. The apparent on–off behavior is observed even at the weak illumination of $P_{\mathrm{in}} = 0.057$ mW/$cm^2$.

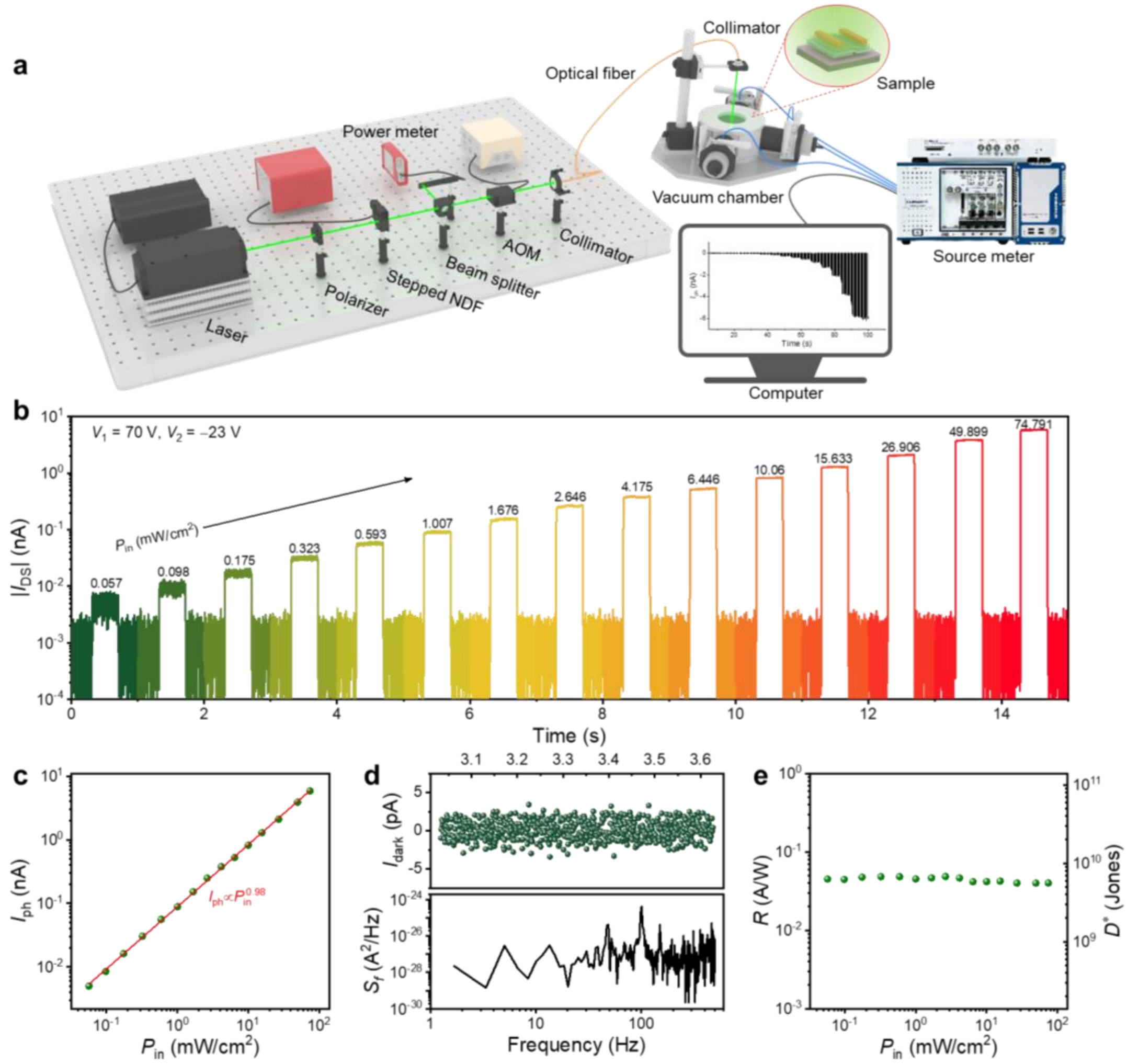

**Fig. S12. Transient photocurrent at p–n state based on one-time test. a)** Schematic of the experimental setup for transient photocurrent measurement. A stepped neutral density filter (NDF) enables power–dependent characterization in a single test. The 532 nm laser is modulated by an acousto–optic modulator (AOM) at 1 Hz with a 40% duty cycle. **b)** Transient current under laser illumination (0.057–74.791 mW/cm$^2$) for the p–n state programmed by a (70, –23) V dipulse, when $V_{\mathrm{DS}} = 0$ V. **c)** Photocurrent extracted from the transient current. **d)** Dark current (upper) and the corresponding calculated noise spectral density (lower). **e)** Calculated responsivity and detectivity as a function of incident laser power density.

The photocurrent ($I_{\mathrm{ph}}$) was extracted from the transient measurements using the relationship:

$$I_{\mathrm{ph}} = I_{\mathrm{illu.}} - I_{\mathrm{dark}}$$

where $I_{\mathrm{illu.}}$ and $I_{\mathrm{dark}}$ denote the current under illumination and dark conditions, respectively. The extracted photocurrent was fitted as $I_{\mathrm{ph}} \propto P_{\mathrm{in}}^{0.98}$ (Fig. S12c), demonstrating near–perfect linearity with incident laser power density, consistent with the characteristic response of a conventional p–n junction. The responsivity, a key metric of input–output efficiency, is given by

$$R = \frac{I_{\mathrm{ph}}}{P_{\mathrm{in}} \cdot A}$$

with the photoactive area $A$. Taking the entire $MoTe_2$ channel area (i.e., $A = 194.7$ μm$^2$) yields the responsivity values plotted in Fig. S12e, with an average of 44.5 mA/W. Besides, the fundamental sensitivity limit is determined by the device noise. The noise spectral density ($S_f$) was calculated via Fourier transform of the dark current, as shown in Fig. S12d, giving $S_{f=1\,\mathrm{Hz}} \approx 1 \times 10^{-28}$ A$^2$/Hz. The noise equivalent power (NEP), defined as

$$NEP = \frac{i_n}{R}$$

in units of W·Hz$^{-1/2}$, represents the minimum detectable light power, where the noise current $i_n = \sqrt{\frac{1}{\Delta f}\int_0^{\Delta f} S_f df}$. For our measurement bandwidth $\Delta f = 1$ Hz, the average NEP is approximately 0.23 pW·Hz$^{-1/2}$. The sensitivity is also quantified by the normalized signal-to-noise ratio over the active area and bandwidth (i.e., detectivity):

$$D^* = \frac{\sqrt{A}}{NEP}$$

in units of cm·Hz$^{1/2}$·W$^{-1}$ or Jones, which can be further simplified to $D^* = R\sqrt{A/S_f}$ for a small measuring bandwidth. The calculated $D^*$ of our device is shown in Fig. S12e, with an average value of $6.2 \times 10^9$ Jones.

A parallel analysis was performed for the n–p configuration programmed by a (–60, 24) V dipulse, are shown in Fig. S13. The photocurrent exhibits a linear dependence on incident laser power density, scaling as $I_{\mathrm{ph}} \propto P_{\mathrm{in}}^{0.96}$, although with minor saturation observed at the highest illumination intensities. The noise spectral density is also obtained through Fourier transform of the current–time data under dark conditions (Fig. S13c), and $S_{f=1\,\mathrm{Hz}} \sim 1 \times 10^{-28}$ A$^2$/Hz. The resulting average responsivity, NEP and detectivity of the device are approximated as 36.2 mA/W, 0.28 pW·Hz$^{-1/2}$ and $5.1 \times 10^9$ Jones, respectively, as shown in Fig. S13d.

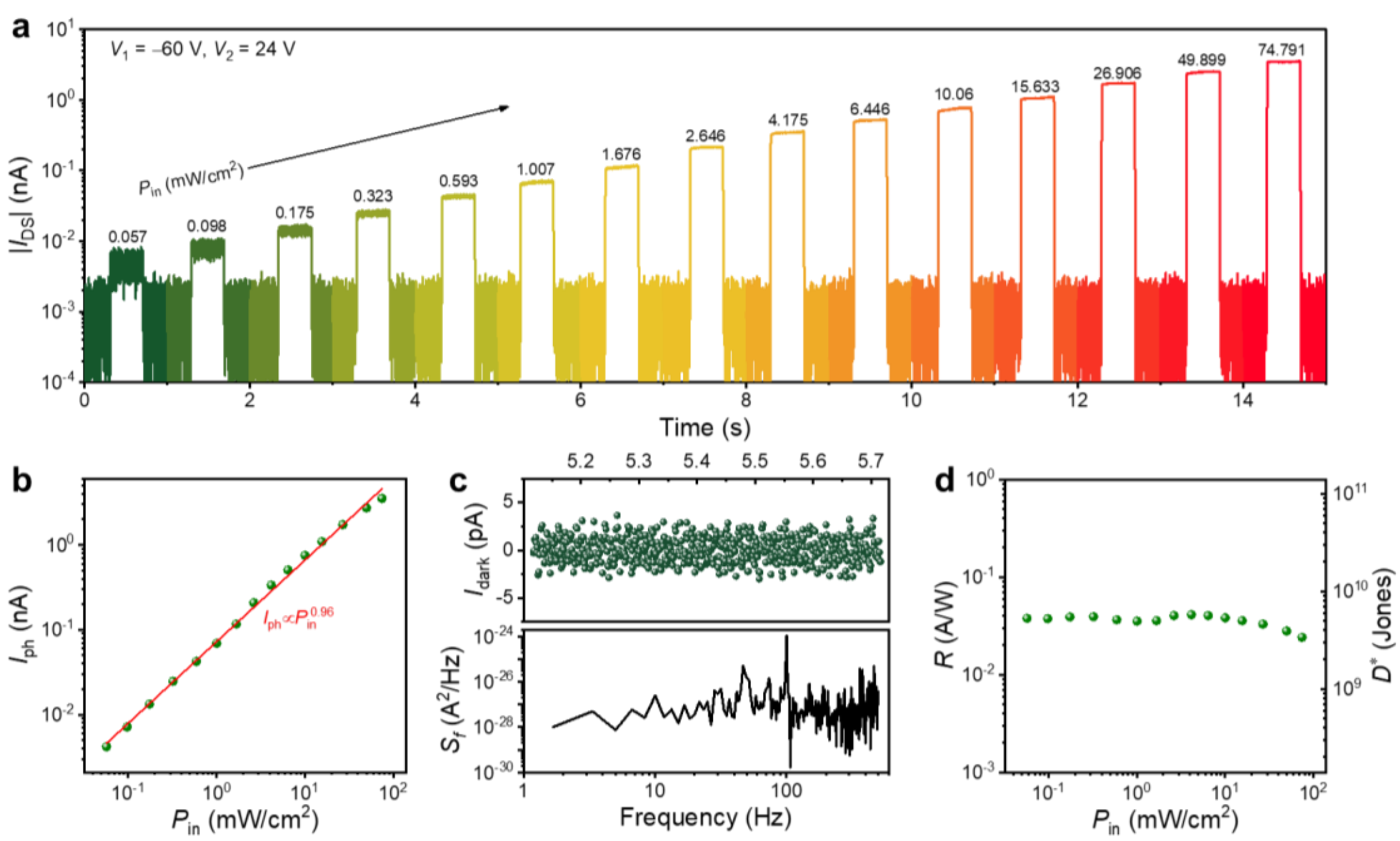

**Fig. S13. Transient photocurrent at n–p state based on one-time test**. **a)** Transient current under laser illumination (0.057–74.791 mW/cm$^2$) for the n–p state programmed by a (–60, 24) V dipulse. **b)** Extracted photocurrent from the transient current. **c)** Dark current (upper) and the calculated noise spectral density (lower) of the device. **d)** Calculated responsivity and detectivity versus the laser power density.

## S7. In-sensor image processing and action classification

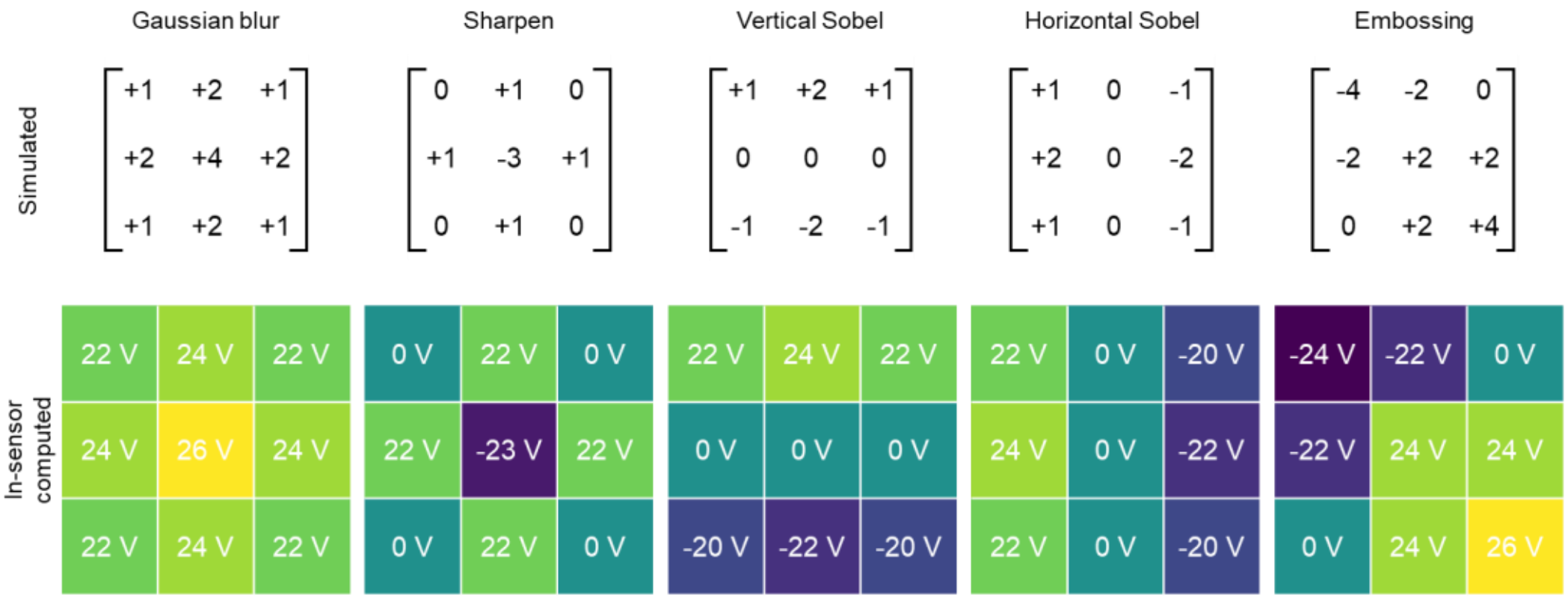

**Fig. S14. Kernel operators for image processing.** Upper row: kernel matrices for conventional software–based processing; lower row: corresponding programming voltage matrices of the following pulse $V_2$ for in–sensor hardware implementation. For all operations, the initial pulse is fixed at $V_1 = \pm 70$ V, with the polarity being opposite to $V_2$, and $V_1 = 70$ V when $V_2 = 0$.

By using different 3×3 kernel operators defined in Fig. S14, the input image was processed via both conventional digital convolution and in–sensor hardware implementation. During the processing, each kernel is convolved across the entire input image (accompanied with each scanning 3×3 image patch) to compute the final output. However, for the Sobel operation, the final output ($Y$) is derived by combing the results from the separate vertical ($Y_y$) and horizontal ($Y_x$) Sobel kernels, calculated as:

$$Y = \sqrt{Y_x^2 + Y_y^2}$$

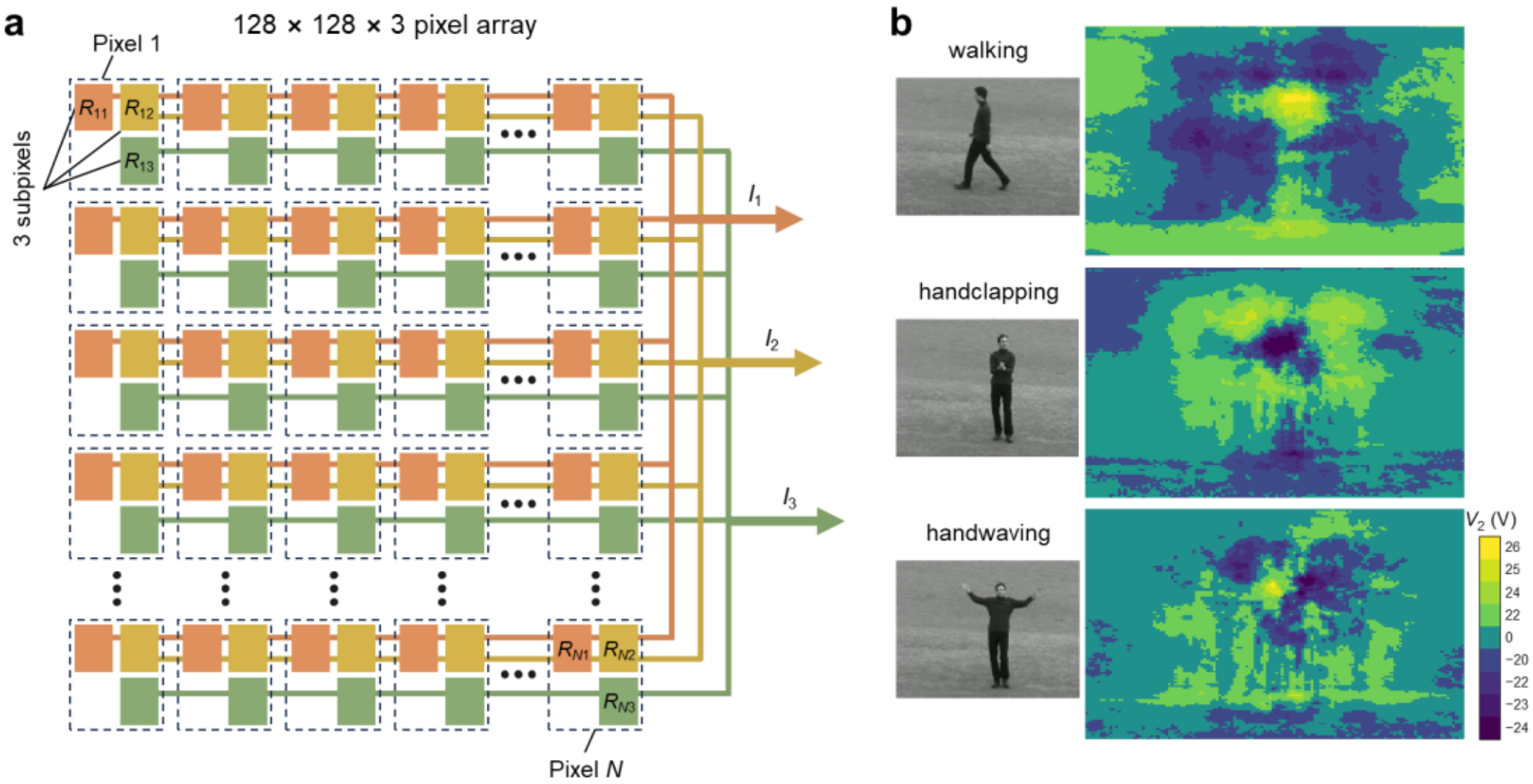

**Fig. S15. Illustration of the in–sensor ANN pixel array. a)** Schematic of the pixel array, comprising $N = 128 \times 128$ pixels. Each pixel integrates three independently programmable subpixels. Subpixels of the same channel are interconnected, generating three distinct and accumulated photocurrent outputs ($I_1$, $I_2$ and $I_3$). **b)** Three target actions of walking, handclapping and handwaving (left) and the corresponding programming gate–voltage ($V_2$) maps for the action recognition (right).

The in–sensor ANN network was simulated using the PyTorch framework. The network architecture comprises a 128×128 input layer (representing the pixel array) and an output layer with three nodes for three–class classification. This process directly realizes the matrix–vector multiplication between input light power vector ($P_N$) and the programmable synaptic weight matrix ($R_{MN}$), yielding a total photocurrent for each output class $M$ governed by:

$$I_M = \sum_{N=1}^{128} R_{MN} \times P_N$$

as illustrated in Fig. S15a.

The training and testing dataset was constructed from three categories (“walking”, “handclapping”, and “handwaving”) in the KTH Action Dataset. During preprocessing, multiple frames were sampled at fixed intervals from each action video and synthesized into a single representative image capturing the core action features for network input. The ANN network was trained offline in the digital domain using a quantization–aware training (QAT) strategy, which explicitly incorporated the physical constraints of the DFG device by restricting synaptic weights to its nine discrete, normalized photoresponse levels. To handle the non–differentiable operations introduced by the discrete constraints during backpropagation, a straight–through estimator (STE) was employed to approximate the gradients. The output layer utilized a softmax activation function, and the network was optimized using a cross–entropy loss function.

In the ANN implementation, the offline–trained synaptic weights were quantized and mapped onto the nine discrete, nonvolatile photoresponse states of the DFG device. Each state was programmed by applying a specific gate–voltage pair (Fig. 4c), thereby physically encoding the trained weights onto the sensor array for in–sensor inference. Accordingly, each of the three 128×128 subpixel arrays was configured with a class–specific spatial map of the programming gate–voltage pair (with the gate–voltage initial pulse $V_1 = -70$ or 70 V and the following pulse $V_2$ maps being shown in Fig. S15b), enabling optimal recognition of the target action.

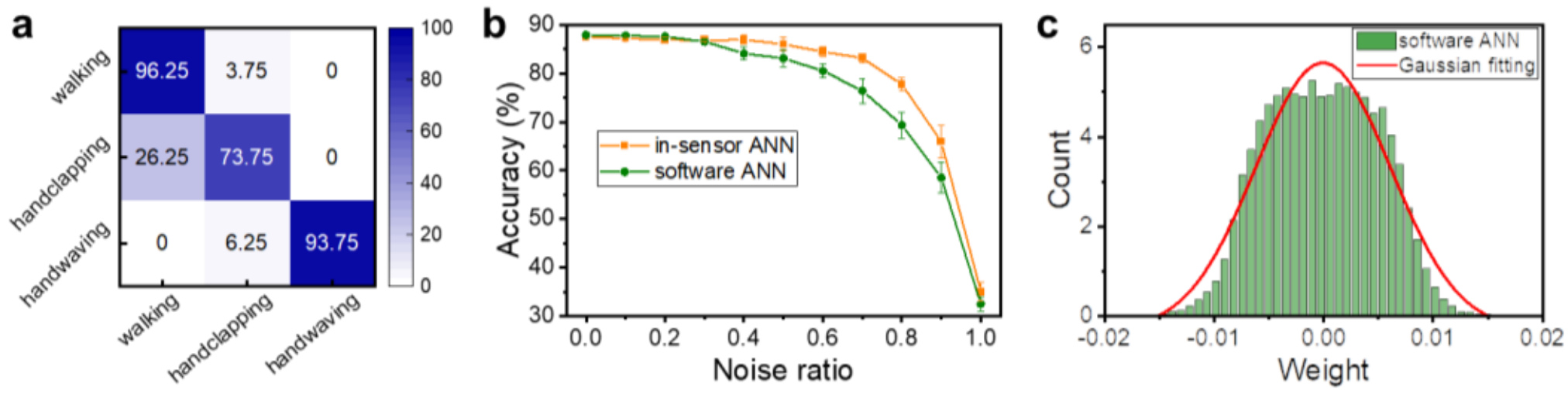

**Fig. S16. Action classification performance of a full-precision software ANN. a)** Confusion matrix of the three–action classification results on the test dataset. **b)** Classification accuracy versus input noise ratio, comparing the full–precision software ANN with the in–sensor ANN. **c)** Distribution of trained synaptic weights for the software ANN.

To validate our physical computing approach, we benchmarked the in–sensor ANN against a full–precision software neural network with identical architecture. The minimal accuracy gap between the in–sensor ANN (87.50%) and the software baseline (87.92%) confirms that the DFG devices effectively execute neuromorphic computation despite their physical constraints. When subjected to salt–and–pepper noise, both the in–sensor and full–precision software models demonstrate exceptional noise immunity (Fig. S16b), which is attributed to their shared weight characteristics. As evidenced by their respective weight distributions, both models exhibit a high degree of effective sparsity, with values overwhelmingly concentrated near or at zero (software: Gaussian–like distribution centered near zero, Fig. S16c; hardware: ~52% of weights at the discrete zero–states, Fig. 5d). This sparsity serves as a natural noise filter during vector–matrix multiplication, nullifying a large fraction of noisy input pixels. Notably, the in–sensor ANN surpasses the software model in noise resilience. This enhanced robustness originates from the discrete, nonvolatile nature of the hardware states. Unlike the continuous, perturbation–sensitive software weights, hardware weights are mapped onto distinct, well–separated levels. The substantial gaps between these discrete physical states provide an inherent quantization margin that effectively suppresses stochastic fluctuations and prevents state–flipping, thus ensuring reliable operation even in high–noise environments.

## S8. Universality verifying of reconfiguration mechanism with ambipolar $WSe_2$

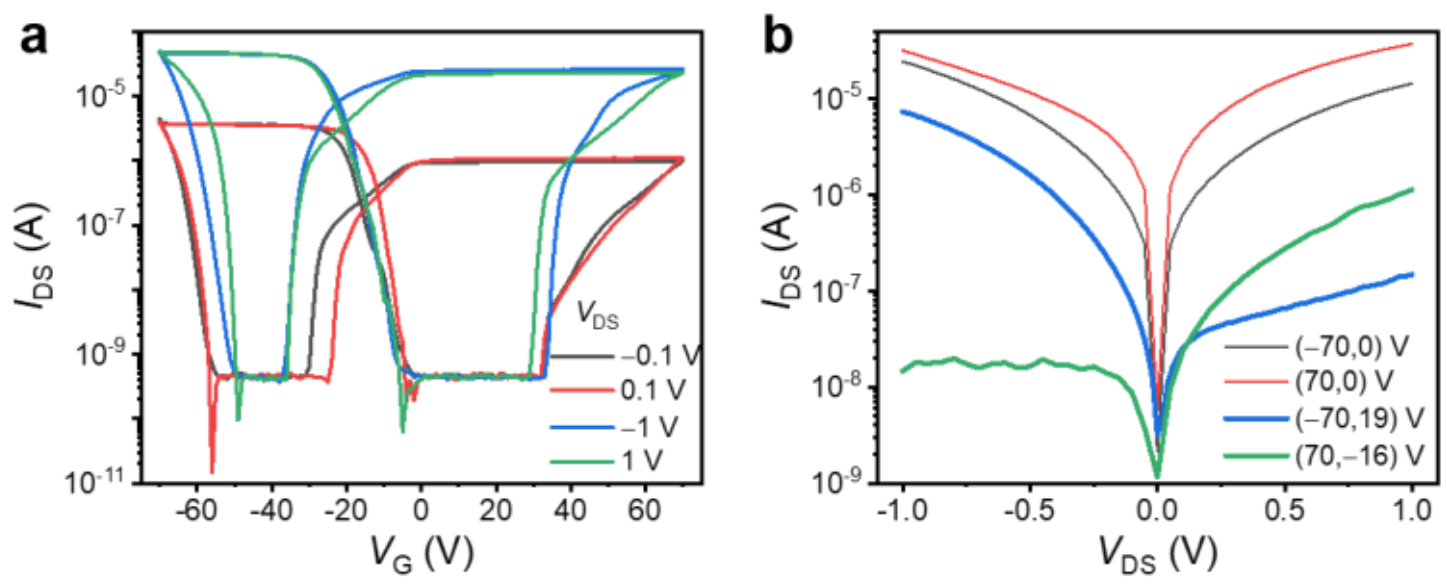

**Fig. S17. Reconfiguration of a $WSe_2$/hBN/Gr DFG device. a)** Dual–sweep transfer curves measured at $V_{DS} = \pm 0.1$ and $\pm 1$ V, respectively. **b)** Output curves following the gate–voltage dipulses of (–70, 0), (70, 0), (–70, 19), and (70, –16) V, respectively.